\documentclass
	[
		10 pt
	]
	{article}
\usepackage
	[
		a4paper,
		left = 2.500 cm,
		right = 2.500 cm,
		top = 2.500 cm,
		bottom = 3.000 cm
	]
	{geometry}
\usepackage{tikz, pgf, pgfplots}
\usetikzlibrary
	{
		3d,
		arrows,											
		arrows.meta,
		automata,										
		backgrounds,
		bending,
		calc,
		chains,
		datavisualization.formats.functions,			
		decorations.pathmorphing,
		decorations.pathreplacing,
		decorations.text,
		fadings,
		fit,
		graphs,
		graphs.standard,
		matrix,
		patterns,
		positioning,									
		quantikz2,
		quotes,
		shadings,
		shadows,
		shadows.blur,
		shapes,
		shapes.geometric,
		trees
	}
\pgfplotsset{compat = newest}
\usepgflibrary
	{
		shapes.misc
	}
\usepgfplotslibrary
	{
		dateplot
	}
\usepackage{tikzpeople}
\usepackage{fontawesome6}
\usepackage{smartdiagram}

\allowdisplaybreaks
\usepackage{amsthm}
\usepackage{amssymb}
\usepackage[bb = boondox]{mathalfa}
\usepackage{dsfont}
\usepackage{newtxmath}
\usepackage{bm}
\usepackage{mathtools}
\usepackage{multicol}
\usepackage{physics2}
\usephysicsmodule{ab.legacy}							
\usephysicsmodule{braket}								
\usephysicsmodule{diagmat}								
\usephysicsmodule{nabla.legacy}							
\usepackage{empheq}
\usepackage[ compat = newest ]{yquant}

\usepackage{sectsty}
\allsectionsfont{\color{WordBlue}}
\usepackage[x11names]{xcolor}
\usepackage{varwidth}
\usepackage{graphicx}
\usepackage{wrapfig}
\usepackage{subcaption}
\usepackage{xurl}                                                                                  %
\usepackage{hyperref}
\hypersetup
	{
		colorlinks,
		linkcolor = {RedPurple},
		citecolor = {WordBlueDarker25},
		urlcolor = {WordAquaDarker50}
	}
\usepackage{tabularray}
\usepackage{booktabs}
\usepackage{colortbl}
\usepackage{float}
\usepackage{nicematrix}
\usepackage{cellspace}
\usepackage{makecell}
\usepackage{multirow}
\usepackage{caption}
\usepackage[ linesnumbered, tworuled, vlined ] {algorithm2e}
\usepackage{appendix}
\usepackage{enumitem}
\usepackage{siunitx}
\usepackage{xintexpr}
\usepackage{spreadtab}
\usepackage{framed}
\usepackage{svg}
\usepackage{lipsum}
\usepackage{listings}
\definecolor{MyLightRed}{RGB}{244, 213, 245}
\definecolor{Purple}{HTML}{911146}
\definecolor{PurpleDark}{RGB}{102, 0, 102}
\definecolor{RedDarkLight}{HTML}{ea005f}
\definecolor{RedDarkLightest}{HTML}{ff0088}
\definecolor{RedPurple}{HTML}{AA007F}
\definecolor{WordPinkAccent1Darker25}{HTML}{B3186D}
\definecolor{WordPinkAccent1Darker50}{HTML}{781049}
\definecolor{WordPinkAccent1Lighter40}{HTML}{EE80BC}
\definecolor{WordPinkAccent1Lighter60}{HTML}{F3AAD2}
\definecolor{WordPinkAccent1Lighter80}{HTML}{F9D4E8}
\definecolor{WordRed}{RGB}{255, 0, 102}
\definecolor{WordRedAccent5Lighter60}{HTML}{F5B5A7}
\definecolor{WordRedAccent5Darker25}{HTML}{B23214}
\definecolor{GreenDark}{HTML}{225522}
\definecolor{GreenLighter1}{HTML}{00B383}
\definecolor{GreenLighter2}{HTML}{00AA7F}
\definecolor{GreenLightest}{HTML}{00FFA0}
\definecolor{GreenTeal}{HTML}{008080}
\definecolor{WordLightGreen}{RGB}{140, 214, 192}
\definecolor{WordGreen}{RGB}{0, 176, 80}
\definecolor{BlueVeryDark}{HTML}{222255}
\definecolor{MyBlue}{RGB}{0, 64, 128}
\definecolor{MyDarkBlue}{RGB}{0, 51, 102}
\definecolor{MyVeryLightBlue}{RGB}{211, 245, 247}
\definecolor{WordBlue}{RGB}{19, 65, 99}
\definecolor{WordBlueDark}{RGB}{46, 116, 181}
\definecolor{WordBlueDarker}{RGB}{31, 78, 121}
\definecolor{WordBlueDarker25}{RGB}{54, 96, 146}
\definecolor{WordBlueDarker50}{RGB}{36, 64, 98}
\definecolor{WordBlueDarkest}{RGB}{0, 32, 96}
\definecolor{WordBlueLight}{RGB}{0, 112, 192}
\definecolor{WordBlueVeryLight}{HTML}{00B0F0}
\definecolor{WordIceBlue}{RGB}{223, 227, 229}
\definecolor{MagentaDark}{RGB}{106, 65, 152}
\definecolor{MagentaLight}{RGB}{128, 100, 162}
\definecolor{MagentaLighter}{RGB}{161, 106, 221}
\definecolor{MagentaVeryDark}{RGB}{97, 75, 128}
\definecolor{MagentaVeryLight}{RGB}{178, 162, 201}
\definecolor{WordAquaAccent1Darker25}{HTML}{276E8B}
\definecolor{WordAquaAccent1Darker50}{HTML}{1A495D}
\definecolor{WordAquaAccent1Lighter40}{HTML}{7FC0DB}
\definecolor{WordAquaAccent1Lighter60}{HTML}{A9D5E7}
\definecolor{WordAquaAccent1Lighter80}{HTML}{D4EAF3}
\definecolor{WordAquaAccent2Darker25}{HTML}{398E98}
\definecolor{WordAquaAccent2Darker50}{HTML}{265F65}
\definecolor{WordAquaAccent2Lighter40}{HTML}{9AD3D9}
\definecolor{WordAquaAccent2Lighter60}{HTML}{BCE1E5}
\definecolor{WordAquaAccent2Lighter80}{HTML}{DDF0F2}
\definecolor{WordAquaDarker25}{HTML}{31869B}
\definecolor{WordAquaDarker50}{HTML}{215967}
\definecolor{WordAquaLighter40}{HTML}{92CDDC}
\definecolor{WordAquaLighter60}{HTML}{B7DEE8}
\definecolor{WordAquaLighter80}{HTML}{DAEEF3}
\definecolor{WordDarkerTeal}{RGB}{48, 82, 80}
\definecolor{WordDarkTeal}{RGB}{72, 123, 119}
\definecolor{WordDarkTealLighter80}{RGB}{207, 223, 234}
\definecolor{WordLightTeal}{RGB}{160, 199, 197}
\definecolor{WordVeryLightTeal}{RGB}{223, 236, 235}
\definecolor{WordTurquoiseLighter80}{RGB}{209, 238, 249}
\definecolor{Brown}{HTML}{666633}
\definecolor{WordGoldAccent1Darker25}{HTML}{C49A00}
\definecolor{WordGoldAccent1Lighter40}{HTML}{FFDF6A}
\definecolor{WordOrangeAccent2Lighter60}{HTML}{FCD3A4}
\definecolor{WordOrangeAccent4Lighter60}{HTML}{F7C5A1}
\definecolor{LavenderBlush}{RGB}{255, 240, 245}
\definecolor{MediumTurquoise}{RGB}{72, 209, 204}
\definecolor{PowderBlue}{RGB}{176, 224, 230}
\definecolor{SkyBlue}{RGB}{135, 206, 235}
\definecolor{Azure2}{RGB}{224, 238, 238}
\definecolor{Azure3}{RGB}{193, 205, 205}
\definecolor{CadetBlue4}{RGB}{83, 134, 139}
\definecolor{DarkSeaGreen1}{RGB}{193, 255, 193}
\definecolor{DeepPink4}{RGB}{139, 10, 80}
\definecolor{Honeydew2}{RGB}{224, 238, 224}
\definecolor{LightSkyBlue1}{RGB}{176, 226, 255}
\definecolor{LightSkyBlue3}{RGB}{141, 182, 205}
\definecolor{LightSkyBlue4}{RGB}{96, 123, 139}
\definecolor{LightSteelBlue1}{RGB}{202, 225, 255}
\definecolor{LightSteelBlue4}{RGB}{110, 123, 139}
\definecolor{MediumPurple1}{RGB}{171, 130, 255}
\definecolor{PaleTurquoise3}{RGB}{150, 205, 205}
\definecolor{PaleVioletRed3}{RGB}{205, 104, 137}
\definecolor{Purple1}{RGB}{155, 48, 255}
\definecolor{SeaGreen1}{RGB}{84, 255, 159}
\definecolor{SeaGreen2}{RGB}{78, 238, 148}
\definecolor{SeaGreen3}{RGB}{67, 205, 128}
\definecolor{SkyBlue1}{HTML}{87CEFF}
\definecolor{SkyBlue4}{RGB}{74, 112, 139}
\definecolor{SteelBlue1}{RGB}{99, 184, 255}
\definecolor{Thistle3}{RGB}{205, 181, 205}
\definecolor{Turquoise4}{RGB}{0, 134, 139}
\definecolor{VioletRed1}{RGB}{255, 62, 150}
\definecolor{VioletRed2}{RGB}{208, 32, 144}
\definecolor{VioletRed3}{RGB}{199, 21, 133}
\definecolor{VioletRed4}{RGB}{139, 10, 80}
\usepackage
	[
		most
	]
	{tcolorbox}
\newcounter{MyCorollary}[section]
\renewcommand{\theMyCorollary}{\thesection.\arabic{MyCorollary}}
\newtcolorbox{corollary} [ 1 ] [ ]
	{
		breakable,
		enhanced,
		enhanced jigsaw,
		skin = enhanced,
		attach boxed title to top left = { xshift = -5.000 mm, yshift = 0.000 mm },
		boxed title style = { boxrule = 0.000 pt, sharp corners = all },
		colbacktitle = RedPurple!90!black,
		coltitle = white,
		fonttitle = \bfseries,
		varwidth boxed title,
		colback = RedPurple!03,
		colframe = RedPurple,
		sharp corners = all,
		toprule = 0.000 mm,
		bottomrule = 0.500 mm,
		leftrule = 0.500 mm,
		rightrule = 0.000 mm,
		code = { \refstepcounter{MyCorollary} },
		title = {Corollary~\theMyCorollary:\if\relax\detokenize{#1}\relax\else~#1\fi},
	}
\newtcolorbox{corollary*} [ 1 ] [ ]
	{
		breakable,
		enhanced,
		enhanced jigsaw,
		skin = enhanced,
		attach boxed title to top left = { xshift = -5.000 mm, yshift = 0.000 mm },
		boxed title style = { boxrule = 0.000 pt, sharp corners = all },
		colbacktitle = RedPurple!90!black,
		coltitle = white,
		fonttitle = \bfseries,
		varwidth boxed title,
		colback = RedPurple!03,
		colframe = RedPurple,
		sharp corners = all,
		toprule = 0.000 mm,
		bottomrule = 0.500 mm,
		leftrule = 0.500 mm,
		rightrule = 0.000 mm,
		title = {Corollary\if\relax\detokenize{#1}\relax\else~#1\fi},
	}
\newcounter{MyDefinition}[section]
\renewcommand{\theMyDefinition}{\thesection.\arabic{MyDefinition}}
\newtcolorbox{definition} [ 1 ] [ ]
	{
		breakable,
		enhanced,
		enhanced jigsaw,
		skin = enhanced,
		attach boxed title to top left = { xshift = -5.000 mm, yshift = 0.000 mm },
		boxed title style = { boxrule = 0.000 pt, sharp corners = all },
		colbacktitle = SkyBlue!70,
		coltitle = SkyBlue!30!black,
		fonttitle = \bfseries,
		varwidth boxed title,
		colback = SkyBlue!15,
		colframe = SkyBlue,
		sharp corners = all,
		toprule = 0.000 mm,
		bottomrule = 0.500 mm,
		leftrule = 0.500 mm,
		rightrule = 0.000 mm,
		code = { \refstepcounter{MyDefinition} },
		title = {Definition~\theMyDefinition:\if\relax\detokenize{#1}\relax\else~#1\fi},
	}
\newtcolorbox{definition*} [ 1 ] [ ]
	{
		breakable,
		enhanced,
		enhanced jigsaw,
		skin = enhanced,
		attach boxed title to top left = { xshift = -5.000 mm, yshift = 0.000 mm },
		boxed title style = { boxrule = 0.000 pt, sharp corners = all },
		colbacktitle = SkyBlue!70,
		coltitle = SkyBlue!30!black,
		fonttitle = \bfseries,
		varwidth boxed title,
		colback = SkyBlue!15,
		colframe = SkyBlue,
		sharp corners = all,
		toprule = 0.000 mm,
		bottomrule = 0.500 mm,
		leftrule = 0.500 mm,
		rightrule = 0.000 mm,
		title = {Definition\if\relax\detokenize{#1}\relax\else~#1\fi},
	}
\newcounter{MyExample}[section]
\renewcommand{\theMyExample}{\thesection.\arabic{MyExample}}
\newtcolorbox{example} [ 1 ] [ ]
	{
		breakable,
		enhanced,
		enhanced jigsaw,
		skin = enhanced,
		attach boxed title to top left = { xshift = -5.000 mm, yshift = 0.000 mm },
		boxed title style = { boxrule = 0.000 pt, sharp corners = all },
		colbacktitle = WordAquaAccent1Darker25,
		coltitle = white,
		fonttitle = \bfseries,
		varwidth boxed title,
		colback = WordAquaAccent1Lighter80!25,
		colframe = WordAquaAccent1Darker25,
		sharp corners = all,
		toprule = 0.000 mm,
		bottomrule = 0.500 mm,
		leftrule = 0.500 mm,
		rightrule = 0.000 mm,
		code = { \refstepcounter{MyExample} },
		title = {Example~\theMyExample:\if\relax\detokenize{#1}\relax\else~#1\fi},
	}
\newtcolorbox{example*} [ 1 ] [ ]
	{
		breakable,
		enhanced,
		enhanced jigsaw,
		skin = enhanced,
		attach boxed title to top left = { xshift = -5.000 mm, yshift = 0.000 mm },
		boxed title style = { boxrule = 0.000 pt, sharp corners = all },
		colbacktitle = WordAquaAccent1Darker25,
		coltitle = white,
		fonttitle = \bfseries,
		varwidth boxed title,
		colback = WordAquaAccent1Lighter80!25,
		colframe = WordAquaAccent1Darker25,
		sharp corners = all,
		toprule = 0.000 mm,
		bottomrule = 0.500 mm,
		leftrule = 0.500 mm,
		rightrule = 0.000 mm,
		title = {Example\if\relax\detokenize{#1}\relax\else~#1\fi},
	}
\newcounter{MyLemma}[section]
\renewcommand{\theMyLemma}{\thesection.\arabic{MyLemma}}
\newtcolorbox{lemma} [ 1 ] [ ]
	{
		breakable,
		enhanced,
		enhanced jigsaw,
		skin = enhanced,
		attach boxed title to top left = { xshift = -5.000 mm, yshift = 0.000 mm },
		boxed title style = { boxrule = 0.000 pt, sharp corners = all },
		colbacktitle = PaleVioletRed3!50,
		coltitle = black,
		fonttitle = \bfseries,
		varwidth boxed title,
		colback = WordPinkAccent1Lighter80!12,
		colframe = WordPinkAccent1Darker50,
		sharp corners = all,
		toprule = 0.000 mm,
		bottomrule = 0.500 mm,
		leftrule = 0.500 mm,
		rightrule = 0.000 mm,
		code = { \refstepcounter{MyLemma} },
		title = {Lemma~\the\theMyLemma:\if\relax\detokenize{#1}\relax\else~#1\fi},
	}
\newtcolorbox{lemma*} [ 1 ] [ ]
	{
		breakable,
		enhanced,
		enhanced jigsaw,
		skin = enhanced,
		attach boxed title to top left = { xshift = -5.000 mm, yshift = 0.000 mm },
		boxed title style = { boxrule = 0.000 pt, sharp corners = all },
		colbacktitle = PaleVioletRed3!50,
		coltitle = black,
		fonttitle = \bfseries,
		varwidth boxed title,
		colback = WordPinkAccent1Lighter80!12,
		colframe = WordPinkAccent1Darker50,
		sharp corners = all,
		toprule = 0.000 mm,
		bottomrule = 0.500 mm,
		leftrule = 0.500 mm,
		rightrule = 0.000 mm,
		title = {Lemma\if\relax\detokenize{#1}\relax\else~#1\fi},
	}
\newcounter{MyProposition}[section]
\renewcommand{\theMyProposition}{\thesection.\arabic{MyProposition}}
\newtcolorbox{proposition} [ 1 ] [ ]
	{
		breakable,
		enhanced,
		enhanced jigsaw,
		skin = enhanced,
		attach boxed title to top left = { xshift = -5.000 mm, yshift = 0.000 mm },
		boxed title style = { boxrule = 0.000 pt, sharp corners = all },
		colbacktitle = cyan7!50,
		coltitle = black,
		fonttitle = \bfseries,
		varwidth boxed title,
		colback = cyan9!25,
		colframe = cyan5,
		sharp corners = all,
		toprule = 0.000 mm,
		bottomrule = 0.500 mm,
		leftrule = 0.500 mm,
		rightrule = 0.000 mm,
		code = { \refstepcounter{MyProposition} },
		title = {Proposition~\theMyProposition:\if\relax\detokenize{#1}\relax\else~#1\fi},
	}
\newtcolorbox{proposition*} [ 1 ] [ ]
	{
		breakable,
		enhanced,
		enhanced jigsaw,
		skin = enhanced,
		attach boxed title to top left = { xshift = -5.000 mm, yshift = 0.000 mm },
		boxed title style = { boxrule = 0.000 pt, sharp corners = all },
		colbacktitle = cyan7!50,
		coltitle = black,
		fonttitle = \bfseries,
		varwidth boxed title,
		colback = cyan9!25,
		colframe = cyan5,
		sharp corners = all,
		toprule = 0.000 mm,
		bottomrule = 0.500 mm,
		leftrule = 0.500 mm,
		rightrule = 0.000 mm,
		title = {Proposition\if\relax\detokenize{#1}\relax\else~#1\fi},
	}
\newcounter{MyTheorem}[section]
\renewcommand{\theMyTheorem}{\thesection.\arabic{MyTheorem}}
\newtcolorbox{theorem} [ 1 ] [ ]
	{
		breakable,
		enhanced,
		enhanced jigsaw,
		skin = enhanced,
		attach boxed title to top left = { xshift = -5.000 mm, yshift = 0.000 mm },
		boxed title style = { boxrule = 0.000 pt, sharp corners = all },
		colbacktitle = WordAquaAccent2Darker25,
		coltitle = white,
		fonttitle = \bfseries,
		varwidth boxed title,
		colback = WordAquaAccent2Lighter80,
		colframe = WordAquaAccent2Darker25,
		sharp corners = all,
		toprule = 0.000 mm,
		bottomrule = 0.500 mm,
		leftrule = 0.500 mm,
		rightrule = 0.000 mm,
		code = { \refstepcounter{MyTheorem} },
		title = {Theorem~\theMyTheorem:\if\relax\detokenize{#1}\relax\else~#1\fi},
	}
\newtcolorbox{theorem*} [ 1 ] [ ]
	{
		breakable,
		enhanced,
		enhanced jigsaw,
		skin = enhanced,
		attach boxed title to top left = { xshift = -5.000 mm, yshift = 0.000 mm },
		boxed title style = { boxrule = 0.000 pt, sharp corners = all },
		colbacktitle = WordAquaAccent2Darker25,
		coltitle = white,
		fonttitle = \bfseries,
		varwidth boxed title,
		colback = WordAquaAccent2Lighter80,
		colframe = WordAquaAccent2Darker25,
		sharp corners = all,
		toprule = 0.000 mm,
		bottomrule = 0.500 mm,
		leftrule = 0.500 mm,
		rightrule = 0.000 mm,
		title = {Theorem\if\relax\detokenize{#1}\relax\else~#1\fi},
	}
\newcounter{mathseed}
\pgfmathsetseed{\arabic{mathseed}}
\pgfdeclarelayer{background}
\pgfsetlayers{background, main}
\pgfdeclaredecoration{irregular fractal line}{init}
{
	\state{init}[width=\pgfdecoratedinputsegmentremainingdistance]
	{
		\pgfpathlineto{\pgfpoint{random*\pgfdecoratedinputsegmentremainingdistance}{(random*\pgfdecorationsegmentamplitude-0.02)*\pgfdecoratedinputsegmentremainingdistance}}
		\pgfpathlineto{\pgfpoint{\pgfdecoratedinputsegmentremainingdistance}{0pt}}
	}
}
\def\tornpaper#1{%
	\ifthenelse{\isodd{\value{mathseed}}}
	{%
		\tikz
		{
			\node[inner sep = 1em] (A) {#1};		
			\begin{pgfonlayer}{background}			
				\fill[paper]						
				\pgfextra{\pgfmathsetseed{\arabic{mathseed}}\addtocounter{mathseed}{1}}%
				{decorate[irregular cloudy border]{decorate{decorate{decorate{decorate[ragged border]{
										(A.north west) -- (A.north east)
				}}}}}}
				-- (A.south east)
				\pgfextra{\pgfmathsetseed{\arabic{mathseed}}}%
				{decorate[irregular spiky border]{decorate{decorate{decorate{decorate[ragged border]{
										-- (A.south west)
				}}}}}}
				-- (A.north west);
			\end{pgfonlayer}
		}
	}
	{%
		\tikz{
			\node[inner sep=1em] (A) {#1};  
			\begin{pgfonlayer}{background}  
				\fill[paper] 
				\pgfextra{\pgfmathsetseed{\arabic{mathseed}}\addtocounter{mathseed}{1}}%
				{decorate[irregular spiky border]{decorate{decorate{decorate{decorate[ragged border]{
										(A.north east) -- (A.north west)
				}}}}}}
				-- (A.south west)
				\pgfextra{\pgfmathsetseed{\arabic{mathseed}}}%
				{decorate[irregular cloudy border]{decorate{decorate{decorate{decorate[ragged border]{
										-- (A.south east)
				}}}}}}
				-- (A.north east);
		\end{pgfonlayer}}
	}
}
\usepackage [ scale = 2 ] {ccicons}
\usepackage{changepage}
\usepackage{standalone}                                                                            %
\title
	{
		AI-Farol: Co-Evolutionary Dynamics in a Multi-Agent Two-Sided Learning Framework
	}

\usepackage{orcidlink}
\newcommand{\orcidicon}[1]{\href{https://orcid.org/#1}{\includegraphics[height=\fontcharht\font`\B]{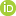}}}

\author
	{
		Iosif Polenakis$^1$\orcidicon{0000-0002-6427-5519},
		Kalliopi Kastampolidou$^2$\orcidicon{0000-0003-3607-9569}
		and
		Theodore Andronikos$^1$\orcidicon{0000-0002-3741-1271}
		\\[10 pt]
		$^1$
		Department of Informatics, Ionian University, \\
		7 Tsirigoti Square, 49100 Corfu, Greece; \\
		\{ipolenakis,andronikos\}@ionio.gr
		\\[8 pt]
		$^2$
		Institute of Applied Biosciences, \\
		Centre for Research and Technology Hellas (CERTH), Thessaloniki, Greece; \\
		kkastampolidou@certh.gr
	}
\begin{document}

\maketitle

\begin{abstract}
	The El Farol Bar game is a classical model of coordination under uncertainty that traditionally treats the venue as a passive constraint. In this work, we reconceptualize the problem by modeling the bar as a strategic player endowed with AI-driven learning capabilities. We extend the original framework in two principal directions: first, by introducing partial observability, whereby agents observe only subsets of past attendees; and second, by transforming the bar from a passive capacity threshold into an active mechanism designer that adjusts pricing policies to balance revenue, utilization, and sustainability constraints. Agents employ AI-based learning to form beliefs and adapt attendance strategies under incomplete information, while the bar applies policy learning to optimize dynamic pricing. The resulting two-sided learning system frames coordination as a co-evolutionary process between boundedly rational agents and an adaptive institution, offering insights into congestion management, resource allocation, and mechanism design in complex adaptive systems.
	\\[12 pt]
\end{abstract}
\section{Introduction} \label{sec: Introduction}

The El Farol Bar problem is a well-known model of coordination under uncertainty in which agents repeatedly decide whether to attend a crowded venue based on limited information about others' behavior~\cite{Arthur1994}. In this work, we extend the classical framework along two dimensions: we introduce \emph{partial observability}, and we model the bar as an active, AI-driven decision maker. Agents employ learning mechanisms to adapt their attendance strategies under incomplete information, while the bar dynamically adjusts its pricing policy to optimize revenue, utilization, and long-run sustainability. This yields a two-sided adaptive system in which both agents and the institution co-evolve through learning and strategic interaction. For a full treatment of partial observability in this setting, we refer the interested reader to \cite{Polenakis2026}. The present work extends the framework further, focusing on the co-evolutionary dynamics between the bar's AI-driven pricing and the agents' learning.

\subsection{Motivation \& related work} \label{sec: Motivation & Related Work}

Arthur's formulation~\cite{Arthur1994} captures a fundamental coordination problem. Each week $N$ agents independently decide whether to attend a bar that is enjoyable only when not overcrowded. Coordination in advance is impossible, and each agent can only observe past attendance records. The problem admits no purely deductive solution; agents must learn~\cite{Fogel1999}. This framework has been applied across traffic routing~\cite{Klugl2004}, bandwidth allocation, load balancing~\cite{Galstyan2005}, and high-frequency trading~\cite{Marsili2000}, and epidemic spread modeling~\cite{Bertolotti2025}, and to the study of how competing algorithms co-adapt for shared resources.

The Minority Game~\cite{Challet1997} strips this resource competition down to a binary choice: at each round, an odd number of agents pick one of two options. Only those on the less-popular side come out ahead. The aggregate outcome is $A(t) = \sum_{ i = 1 }^{ N } a_{ i } ( t )$ with agent payoff $u_i(t) = -a_i(t) \cdot \operatorname{sgn}(A(t))$. System behavior is governed by $\alpha = 2^M/N$, a phase parameter balancing available information against competition, giving rise to market-like statistical properties without explicit design~\cite{Challet2000, Johnson1998}. A related extension moves from a single venue to many: the Kolkata Paise Restaurant (KPR) problem~\cite{Chakrabarti2009, Ghosh2013} places $N$ agents against $N$ restaurants, each able to serve only one customer. Writing $r_i(t)$ for agent $i$'s restaurant choice, occupancy is $n_j(t) = \sum_{i} \mathbb{I}(r_i(t)=j)$, and service is all-or-nothing per restaurant. The natural performance measure is utilization, $f(t) = \frac{1}{N}\sum_j \mathbb{I}(n_j(t)=1)$: agents choosing uniformly at random leave roughly $1/e \approx 0.368$ of capacity served, a figure that learning-based strategies can push up to the $0.5$--$0.8$ range~\cite{Ghosh2010, Biswas2024}. The problem has also found connections to combinatorial optimization~\cite{Kastampolidou2022} and to quantum-theoretic treatments of games~\cite{Ramzan2013}.

Despite their influence, all three frameworks share a fundamental blind spot: venues are passive. Real venues adjust prices, impose cover charges, and manage demand through reservation systems, interventions that alter the very coordination problem customers face. The present work fills this gap because it elevates the bar to a strategic player, having its own objectives, constraints, and learning capabilities, and studies the resulting two-sided adaptive system. Autonomous pricing agents that learn and adapt in competitive markets have been studied outside the El Farol tradition as well~\cite{Greenwald1999}, though not with the venue itself cast as a party to a repeated coordination game with its own customer base. As analyzed in~\cite{Polenakis2026}, partial observability is a key driver of asymmetric coordination failure; here we examine how the bar's dynamic pricing interacts with and shapes the agents' learning dynamics.

This broader perspective supports a game-theoretic treatment in which the bar is itself a strategic player. Beyond these foundational formulations, a substantial body of work has adapted the bar/restaurant paradigm to richer agent architectures and application domains, though largely without granting the venue itself any strategic role. Extending the KPR line, Park and Saad~\cite{Park2017} recast resource allocation among internet-of-things devices with imperfect mutual knowledge as a many-restaurant coordination game, while Ghosh and Chakrabarti~\cite{Ghosh2017} show that reinforcement learning over restricted information sets can substantially raise the utilization rate achievable under such finite-information settings. Agarwal et al.~\cite{Agarwal2016} study a related distributed coordination game with many equally-preferred equilibria, showing that simple reinforcement-based heuristics cause agents to self-organize into clusters, most of which are transitory before the system settles. On the classical single-venue problem, Franke~\cite{Franke2003} was among the first to replace Arthur's inductive forecasters with stimulus-response reinforcement learning, contrasting belief-based and reinforcement-based adaptation as distinct routes to the same coordination dynamics, and Miramontes Hercog and Fogarty~\cite{Hercog2001, Hercog2004} apply co-evolving classifier systems to the El Farol setting, identifying an emergent ``vacillating agent'' that sacrifices its own payoff to stabilize attendance for the rest of the population; a later extension generalizes this mechanism to multiple, simultaneously discoordinating bars for manufacturing scheduling~\cite{Hercog2013}. Oh and Smith~\cite{Oh2008} instead target the tension between selfish and centrally optimal outcomes directly, proposing a social learning scheme that lets self-interested agents cooperate to reduce the price of anarchy while explicitly bounding the communication cost of doing so, with the El Farol bar problem used as a validating testbed. An earlier, related redesign is due to Farago et al.~\cite{Farago2002}, who show that the Nash equilibria of the unmodified Santa Fe bar problem are either unfair or inefficient, and propose a restructured version of the game under which fictitious play, no-regret learning, and Q-learning agents converge to fair and efficient outcomes. More recently, Takata et al.~\cite{Takata2025} replace hand-crafted learning rules with LLM-based agents in a spatially extended bar problem, finding that language models develop spontaneous, only partially rational attendance motivations shaped by pretraining rather than by the formal payoff structure alone, a result that motivates our own use of AI-driven decision-making on the agent side. Separately, Janssen et al.~\cite{Janssen2019} address a complementary methodological gap, proposing causal discovery as a means of explaining \emph{why} emergent regularities arise in agent-based models rather than merely detecting that they do, a diagnostic lens that is largely absent from the El Farol and KPR literatures surveyed above. Across all of these extensions, however, the venue, bar, restaurant, or resource, remains an exogenous capacity constraint rather than an adaptive party to the game.

For completeness we should mention that game-theoretic tools have proven effective in different  unconventional settings, including quantum cryptographic protocols \cite{Bennett1984, Ampatzis2021, Ampatzis2022,  Ampatzis2023, Andronikos2023, Andronikos2023a, Andronikos2023b, Karananou2024, Andronikos2024, Andronikos2024a, Andronikos2024b, Andronikos2025, Andronikos2025c}, quantum classification of functions \cite{Andronikos2025a, Andronikos2025b}. and bio-inspired evolutionary games \cite{Kastampolidou2020, Kastampolidou2021, Kastampolidou2023}. There are situations where quantum games, popularized since 1999 \cite{Meyer1999, Eisert1999}, often outperform classical strategies \cite{Andronikos2018,Andronikos2021,Andronikos2022a}, as exemplified by the Prisoners’ Dilemma \cite{Eisert1999} and other abstract games \cite{Koh2024}. Let us note that in a way quantum games can also prove useful in function classification, especially for functions exhibiting specific properties. In a certain sense, the Deutsch-Jozsa algorithm \cite{Deutsch1992} pioneered this field, followed by extensions such as multidimensional variants \cite{Cleve1998}, generalizations of balanced functions \cite{Chi2001, Holmes2003}, and further extension \cite{Ballhysa2004, Qiu2018, OssorioCastillo2023}. Recent works such as \cite{Andronikos2025a} and \cite{Andronikos2025b} targeted imbalanced functions, proposing the Boolean Quantum Classifiers for classifying functions with specific behavioral patterns. For a very recent probabilistic analysis of the classification outcomes with respect to the Hamming distance see also \cite{Andronikos2026}. Algorithms of the latter category typically have the form of a quantum game featuring Alice and Bob, using the engaging nature of games to clarify complex concepts. Furthermore, many classical systems can be transformed into quantum versions, including political frameworks, as demonstrated in recent studies \cite{Andronikos2022}. Games can be useful also in the study of biological systems, a trend that has gathered momentum during the last decade \cite{Theocharopoulou2019, Kastampolidou2020a, Kostadimas2021}.

\textbf{Contribution.}

This work addresses a longstanding blind spot in the El Farol Bar problem and its relatives (the Minority Game, the Kolkata Paise Restaurant problem): in all of these frameworks the venue itself is treated as a passive constraint, even though real venues actively manage demand through pricing, cover charges, and reservation policies. We reconceptualize the bar as a strategic, AI-driven player in its own right, and combine this with two extensions to the customer side, partial observability of past attendance and AI-based belief formation and strategy adaptation, to obtain a genuinely two-sided, co-evolutionary learning system rather than a one-sided optimization problem against a fixed rule. Specifically, the contribution of this work can be summarized as follows:

\begin{itemize}
	[ left = 0.200 cm, labelsep = 0.400 cm ]
	\item	[\textcolor{WordAquaAccent1Darker25} \faListUl]
	\textbf{Modeling.}
	We generalize El Farol-type games to incorporate mechanism design and two-sided learning, extending and generalizing the direction introduced in~\cite{Polenakis2026}. This contribution is structural and definitional, it specifies the strategy spaces, objectives, and solution concepts of the two-sided game, rather than a set of formal existence, uniqueness, or convergence theorems; establishing such results is flagged as open work in Section \ref{sec: Conclusions}.
	\item	[\textcolor{WordAquaAccent1Darker25} \faListUl]
	\textbf{Analytical.}
	We present a novel characterization of the bar-side and customer-side AI capabilities, demand forecasting, reinforcement-learning-based dynamic pricing, Bayesian belief updating, and no-regret learning, and of the co-evolutionary dynamics that emerge when both sides adapt simultaneously.
	\item	[\textcolor{WordAquaAccent1Darker25} \faListUl]
	\textbf{Methodological.}
	We outline an adaptation of approximate Bayesian Nash equilibrium, coarse correlated equilibrium, and Stackelberg equilibrium to this jointly-adaptive setting, together with a welfare framework that separates coordination efficiency from one-sided surplus extraction.
	\item	[\textcolor{WordAquaAccent1Darker25} \faListUl]
	\textbf{Practical.}
	We perform an extensive simulation study across sixteen agent–bar pairings and three time horizons that empirically tests these concepts, including a direct check of the no-regret/CCE claim against simulated play and a diagnosis of which observed instabilities are genuinely co-evolutionary versus attributable to demand-model misspecification.
\end{itemize}

\subsection*{Organization} \label{subsec: Organization}

Section \ref{sec: The Bar As A Strategic Player} formalizes the bar as a strategic player: its strategy space, objectives, and the two-sided game. Section \ref{sec: AI-Augmented Bar Intelligence & Co-Evolutionary Dynamics} introduces AI capabilities on both sides and analyzes co-evolutionary dynamics. Section \ref{sec: Equilibrium Analysis & Welfare} addresses equilibrium concepts and welfare analysis. Section \ref{sec: Model Evaluation} tests these ideas in simulation, checking whether the co-evolutionary dynamics and equilibrium concepts of the preceding sections actually show up under the model's own learning rules. Section~\ref{sec: Conclusions} concludes with applications and open questions.

\section{The Bar as a strategic player} \label{sec: The Bar As A Strategic Player}

In this Section, we establish the underlying strategic environment. We first cast the bar itself as an optimizing agent, defining its pricing instrument, its revenue and profit accounting, and the operating constraints it must respect. We then combine this with the customer side to obtain a fully specified two-sided dynamic game, including the period-by-period timing protocol and the resulting hierarchical, Stackelberg-like structure that the equilibrium concepts of Section \ref{sec: Equilibrium Analysis & Welfare} must later accommodate.

\subsection{Strategy space, objectives \& payoffs} \label{subsec: Strategy Space, Objectives & Payoffs}

The bar's principal lever for shaping demand is price. Formally, the effective price at period $t$ is:

\begin{align}
	p_t
	=
	p_0 - d_t
	\ ,
\end{align}

where $p_0$ is a baseline price and $d_t \in [0, d_{ \max }]$ is a discount chosen by the bar. The bar may additionally exercise control over capacity management, information disclosure, and promotional targeting, but we focus on pricing as the main strategic instrument. The bar's gross revenue, net profit, and operational constraints are:

\begin{align}
	R_t
	=
	p_t K_t
	\ ,
	\qquad
	\Pi_t
	=
	p_t K_t - C_B(K_t)
	\ ,
\end{align}

where $K_t$ is attendance and $C_B(K_t)$ denotes attendance-dependent operating costs. Three binding constraints apply:

\begin{align}
	K_t \leq K_{\max}
	\ ,
	\qquad K_t \geq K_{\min}
	\ ,
	\qquad \Pi_t \geq \Pi_{\min}
	\ .
\end{align}

The bar maximizes long-run expected discounted profit:

\begin{align}
	\max_{\{d_t\}} \,\mathbb{E}\!\left[\sum_{t=0}^{\infty} \beta^t \Pi_t \right], \quad \beta \in (0,1).
\end{align}
\subsection{Two-sided game formulation \& mathematical model} \label{subsec: Two-Sided Game Formulation & Mathematical Model}

Neither party in this setting can be analyzed in isolation: customer attendance depends on the price the bar sets, and the bar's pricing depends on how customers are expected to respond. This mutual dependence is what makes the model a genuine two-player dynamic game rather than a single-sided optimization problem against an exogenous price. A population of $n$ boundedly rational agents indexed by $i \in \{1,\dots,n\}$ repeatedly decide whether to attend; let $a_i^t \in \{0,1\}$ denote agent $i$'s decision and $K_t = \sum_i a_i^t$ be total attendance. Each agent derives utility:

\begin{align}
	u_i^t
	=
	a_i^t \cdot \left[U(K_t) - p_t\right]
	\ ,
\end{align}

where $U(K_t)$ is a concave, congestion-sensitive satisfaction function~\cite{Rosenthal1973}. A key feature of the framework, analyzed in depth in~\cite{Polenakis2026}, is \emph{partial observability}: when agent $i$ attends at time $t$, they observe only a subset $S_i^t \subsetneq \{1,\ldots,n\} \setminus \{i\}$ of the other attendees, while agents who stay home receive no information. From these partial observations, each agent maintains a belief state $b_i^t$ over the attendance probabilities of others, updated from the historical sequence $\{S_i^\tau\}_{\tau < t}$. Agent $i$'s strategy maps beliefs and price to an attendance probability:

\begin{align}
	\label{eq:strategy}
	\sigma_i^t
	\colon
	(b_i^t,\, p_t) \to [0,1]
	\ .
\end{align}

The bar's policy maps its aggregate history to a discount level:

\begin{align}
	\pi^t
	\colon
	\bigl(\{K_\tau, R_\tau\}_{\tau < t}\bigr) \to [0,\, d_{\max}]
	\ .
\end{align}

The system evolves sequentially each period as described below.

\begin{enumerate}
	[
		label = \textcolor {WordAquaAccent2Darker25} {\textbf{(\arabic*)}},
		left = 0.000 cm,
		labelsep = 0.400cm
	]
	\item	
	The bar observes its state and chooses $d_t$ via $\pi^t$, setting $p_t = p_0 - d_t$.
	\item	
	Agents observe $p_t$ and form attendance decisions from $b_i^t$ and $\sigma_i^t$.
	\item	
	Attendance realizes as $K_t = \sum_i a_i^t$.
	\item	
	Attending agents receive $u_i^t = a_i^t \cdot [U(K_t) - p_t]$.
	\item	
	The bar receives $\Pi_t = p_t K_t - C_B(K_t)$.
	\item	
	Attending agents observe subsets $S_i^t$ and update beliefs.
	\item	
	The bar observes $K_t$ and $R_t$ and updates its policy.
\end{enumerate}

\begin{figure}[t!]
	\centering
	\includegraphics[width=0.99\textwidth]{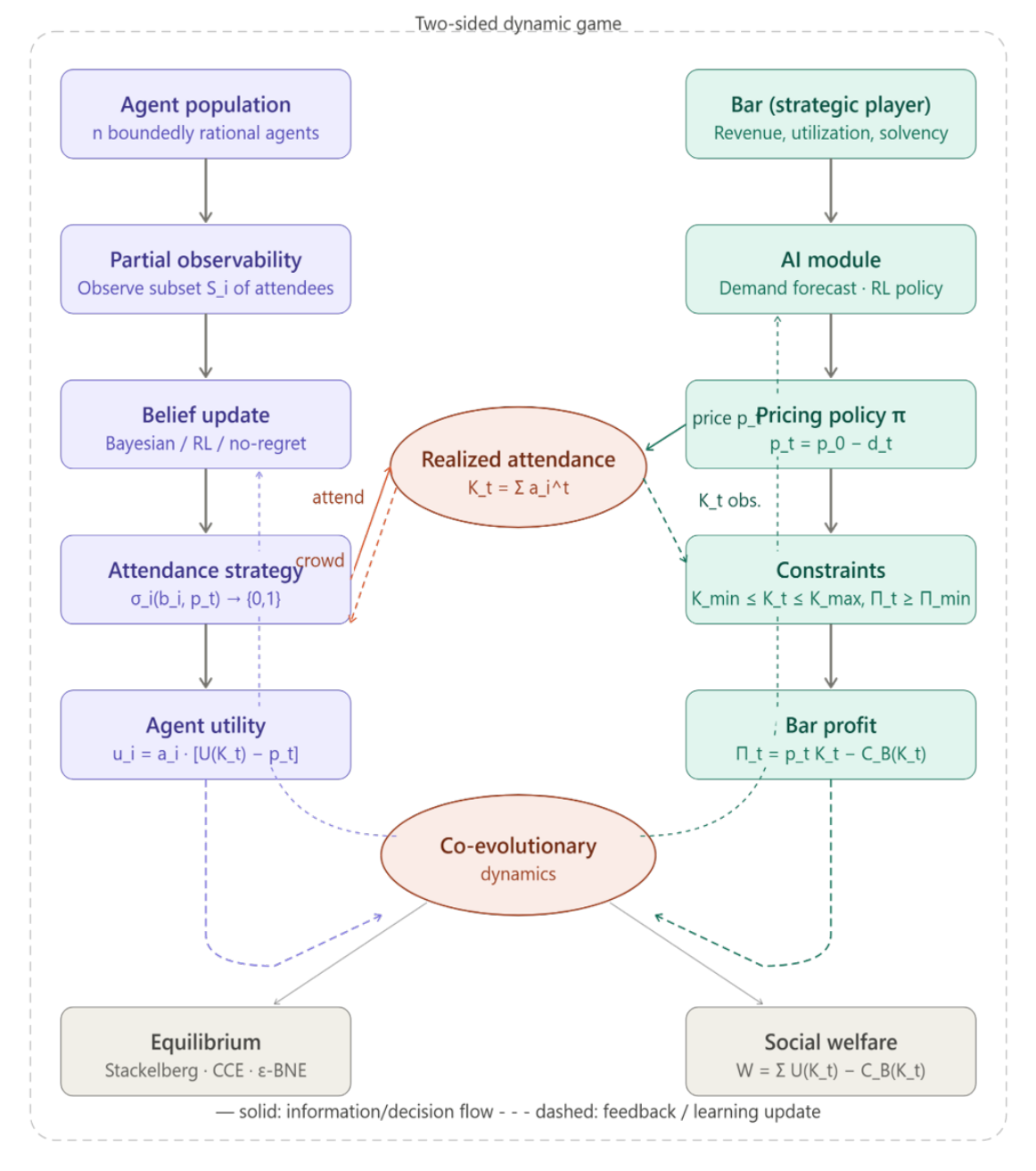}
	\caption{Modified El Farol Bar problem with an AI-driven pricing mechanism. The bar acts as an active player, adjusting prices dynamically based on predictions generated by an AI module trained on past customer behavior, with the goal of optimizing attendance and customer satisfaction.}
	\label{fig: el_farol_ai}
\end{figure}

The game has a hierarchical structure: at the \emph{lower level}, agents choose $\{\sigma_i\}$ given $\pi$ and their beliefs about others; at the \emph{upper level}, the bar selects $\pi$ in anticipation of how agents will respond through their learning dynamics. The equilibrium is not a static Nash equilibrium but a learning-consistent dynamic equilibrium in which both agent strategies and bar policies co-evolve. The architecture of the full AI-Farol framework, comprising the agent side, the bar's AI module, and the co-evolutionary interface, is depicted in Figure~\ref{fig: el_farol_ai}.

\section{AI-augmented Bar intelligence \& co-evolutionary dynamics} \label{sec: AI-Augmented Bar Intelligence & Co-Evolutionary Dynamics}

Having established the strategic environment, next, we proceed by equipping both players with concrete learning algorithms~\cite{zhang2021}. We first describe the statistical and reinforcement-learning tools available to the bar for demand forecasting and dynamic pricing; we then describe the belief-formation and learning rules available to agents operating under the partial observability introduced above; and we close by characterizing the feedback loop that arises once both sides adapt simultaneously, the co-evolutionary dynamic that motivates the equilibrium concepts of Section~\ref{sec: Equilibrium Analysis & Welfare} and is tested against simulation in Section~\ref{sec: Model Evaluation}.

\subsection{Bar-side AI capabilities} \label{subsec: Bar-Side AI Capabilities}

The Bar's optimization problem is genuinely difficult: it must forecast demand from a population of learning, non-stationary agents while observing only aggregate outcomes and operating under binding constraints. AI provides a viable path through these challenges. Throughout, we use ``AI'' as an umbrella term for the specific statistical and reinforcement-learning techniques detailed below, regression and Gaussian-process demand forecasting, tabular reinforcement learning with constraint-penalized rewards, Bayesian belief updating, and no-regret learning, which are established tools in the multi-agent learning and mechanism-design literature rather than a qualitatively new capability; \cite{Franke2003}, for instance, already applies reinforcement learning to the (single-sided) El Farol problem. We retain the term for consistency with the framing adopted in \cite{Polenakis2026}.

\textbf{Demand forecasting.} From observed price--attendance pairs $(p_\tau, K_\tau)_{\tau < t}$, the bar builds an estimate of $P(K_t \mid p_t, \text{history})$. A minimal version of this is a regression on recent lags:

\begin{align}
	\hat{K}_t
	=
	f(p_t,\, K_{t-1},\, \ldots,\, K_{t-M})
	\ ,
\end{align}

where $f(\cdot)$ may be linear, a random forest, or a neural network. A Bayesian approach maintains a posterior over demand parameters, enabling risk-aware decisions that balance exploration and exploitation.

\textbf{Dynamic pricing via reinforcement learning.} Given demand forecasts, the bar solves a Markov decision process~\cite{Bellman1957}, learning a policy $\pi^*: \text{state} \to \text{action}$ that maximizes expected cumulative discounted profit:

\begin{align}
	\pi^*
	=
	\arg\max_\pi \,\mathbb{E}_{\pi}\!\left[\sum_{t=0}^{\infty} \beta^t \Pi_t \right]
	\ .
\end{align}

Operational constraints are incorporated as penalty terms in the reward signal, ensuring the learned policy remains feasible.

\textbf{Learning customer response patterns.} The bar estimates how pricing changes translate into attendance changes:

\begin{align}
	\Delta K_t = g(\Delta p_t,\, \text{history}),
\end{align}

using time-series models such as vector autoregressions to capture lagged effects and feedback dynamics.

\subsection{Customer-side learning} \label{subsec: Customer-Side Learning}

Agents employ AI to navigate their partial observability, as formalized in~\cite{Polenakis2026}. Three mechanisms are available:

\textbf{Belief formation under partial observability.} Using Bayesian inference over observed subsets $\{S_i^\tau\}_{\tau < t}$, agent $i$ estimates the probability that agent $j$ attends:

\begin{align}
	\hat{p}_{ij}^t
	=
	P(a_j^t = 1 \mid \{S_i^\tau\}_{\tau < t})
	\ .
\end{align}

Note that agent $i$ has observations only for periods in which they attended, potentially introducing selection bias that must be accounted for in inference.

\textbf{Strategy adaptation via reinforcement learning.} Agents treat the environment as a Markov game~\cite{Shapley1953} with states encoding beliefs and prices, attend/not-attend actions, and rewards $u_i^t$. Algorithms such as Q-learning~\cite{Watkins1992} or policy gradients improve strategies over time without computing Nash equilibria explicitly~\cite{Franke2003}.

\textbf{No-regret learning.} Rather than committing to a single rule, an agent can hedge across several candidate strategies and reweight them by realized performance, Multiplicative Weights and Follow-the-Regularized-Leader (FTRL) are standard instances~\cite{Auer1995}. Such algorithms carry a guarantee that matters here: over time, the agent does no worse than the single best fixed strategy it could have played in hindsight. The population-level consequence, established by~\cite{Hart2000, Blum2007}, is that if every agent learns this way, the empirical joint distribution of play settles toward a coarse correlated equilibrium, giving a concrete, checkable target for the co-evolutionary dynamics discussed below.

\subsection{Co-evolutionary dynamics} \label{subsec: Co-Evolutionary Dynamics}

When both sides learn simultaneously, mutual feedback loops emerge. Agent strategies evolve in response to the bar's pricing; the bar's policy evolves in response to aggregate agent behavior. Neither side converges in isolation.

Take a bar that has been under-attended for several weeks and responds by cutting its price sharply. Attendance jumps and revenue follows, a short-run win. But the resulting crowd is larger than what customers found comfortable, so satisfaction on that visit drops; customers who had a disappointing night lower their expectation of an uncrowded visit next time, and some stay home. Attendance falls, cutting into revenue, and the bar, reading only the drop in receipts, not the reason behind it, cuts price again to compensate. The two sides end up chasing each other's previous move rather than settling into a stable arrangement. Whether this settles down or keeps oscillating depends on how fast each side adapts: slow learning rates tend toward an approximate equilibrium, while faster mutual adaptation can sustain persistent oscillations or limit-cycle behavior indefinitely. Analyzing these dynamics rigorously requires tools from dynamical systems theory and evolutionary game theory, with mean-field approximations collapsing the heterogeneous customer population into a tractable aggregate.

\section{Equilibrium analysis \& welfare} \label{sec: Equilibrium Analysis & Welfare}

This section investigates the properties exhibited by the two-sided learning system to settle down, and how to judge the outcome once it does. We first adapt three solution concepts, approximate Bayesian Nash equilibrium, coarse correlated equilibrium, and Stackelberg equilibrium~\cite{Simaan1973}, to a setting in which neither side is assumed to compute exact best responses, and we are explicit about which formal guarantees carry over from the stationary, single-shot case and which do not yet have a proof in the jointly-adaptive setting studied here. We then define agent surplus, bar profit, and social welfare, and use these definitions to separate the question of whether the system coordinates efficiently from the separate question of which side captures the resulting surplus, a distinction Section~\ref{sec: Model Evaluation} returns to repeatedly.

\subsection{Equilibrium concepts} \label{subsec: Equilibrium Concepts}

Classical Nash equilibrium~\cite{Nash1951} is ill-suited here due to non-stationarity, incomplete and asymmetric information, and the bar's hierarchical strategic role. We therefore adopt three learning-based solution concepts.

An \emph{approximate Bayesian Nash equilibrium}~\cite{Harsanyi1967} ($\varepsilon$-BNE) obtains when agents best-respond to approximately correct beliefs. Formally, strategy $\sigma_i$ is an $\varepsilon$-best response if:

\begin{align}
	 \label{eq: ebne}
	\mathbb{E}_{a_{-i} \sim b_i}\bigl[u_i(a_i, a_{-i})\bigr] \;\geq\; \max_{a_i'}\, \mathbb{E}_{a_{-i} \sim b_i}\bigl[u_i(a_i', a_{-i})\bigr] - \varepsilon
	\ .
\end{align}

Here $a_i$ is understood to be drawn according to the belief-conditioned strategy map $\sigma_i^t(b_i^t, p_t)$ of Eq.~\eqref{eq:strategy}, so the comparison in Eq.~\eqref{eq: ebne} is over the attendance probabilities that map induces, not over $\sigma_i$ as an abstract object.

A \emph{coarse correlated equilibrium} (CCE) is a weaker, more permissive notion than the correlated equilibrium concept it relaxes~\cite{Aumann1974}: it asks only that switching to some fixed alternative strategy would not have improved an agent's payoff on average over the history of play. This is exactly the guarantee no-regret learning delivers, which is why populations of no-regret learners are the natural empirical route to CCE rather than to Nash equilibrium~\cite{Hart2000}, a contrast documented directly by~\cite{Jafari2001}, who show no-regret dynamics converge to Nash equilibrium only in restricted game classes and can cycle otherwise.

This route is formally established for repeated play of a \emph{fixed} stage game; the no-regret guarantee itself is more robust, holding even against adversarially varying payoffs, but the population-level convergence to a CCE proved in~\cite{Hart2000} is stated for a stationary game. In the present setting the ``stage game'' the agents face is not stationary, since the bar's own pricing policy is simultaneously adapting to aggregate attendance. We do not attempt a formal extension of the Hart--Mas-Colell argument to this jointly-adaptive setting here; instead, Subection \ref{subsec: Checking The No-Regret Claim} checks the CCE signature, vanishing empirical regret, directly against simulated play. That check holds cleanly for three of the four bar types but breaks down, at least within the simulated horizon, against a bar whose own pricing is itself unstable (Section~\ref{subsec: Variability & The Signature Of Persistent Adaptation}). We read this as informative about where the argument is likely to hold rather than as a substitute for a proof that it holds in general.

With the bar as a strategic player, a \emph{Stackelberg equilibrium} is also natural: the bar acts as leader, committing to a pricing policy, while customers respond optimally. This is appropriate when the bar can credibly commit to posted prices. The $\varepsilon$-BNE concept is particularly relevant given the partial observability constraints studied in~\cite{Polenakis2026}, where approximate best-responses arise naturally from limited observations.

In the AI-augmented setting, equilibrium is an emergent property rather than a starting assumption: AI-based learning produces approximately optimal strategies ($\varepsilon$-equilibria), and mutual adaptation may produce persistent fluctuations rather than a fixed point. Beliefs may also diverge from rational expectations due to heterogeneous priors or biased learning algorithms.

\subsection{Welfare \& efficiency analysis} \label{subsec: Welfare & Efficiency Analysis}

To compare outcomes under strategic versus passive bars, we define:

\noindent\textbf{Agent surplus}:

\begin{align}
	W_{\text{agents}}^t
	=
	\sum_{i:\, a_i^t = 1} \bigl[U(K_t) - p_t\bigr]
	\ .
\end{align}

\noindent\textbf{Bar profit}:

\begin{align}
	\Pi_t
	=
	p_t K_t - C_B(K_t)
	\ .
\end{align}

\noindent\textbf{Social welfare}:

\begin{align}
	W_{\text{total}}^t
	=
	W_{\text{agents}}^t + \Pi_t = \sum_{i:\, a_i^t = 1} U(K_t) - C_B(K_t)
	\ .
\end{align}

Notice that price does not appear in $W_{\text{total}}^t$: whatever the bar charges is simply money changing hands between the two sides, so it cannot make the system as a whole better or worse off. What determines efficiency is purely how close realized attendance sits to the level $K^*$ that best trades off enjoyment against operating cost. A bar with pricing power can, in principle, nudge attendance toward $K^*$, but it can just as easily use that power to extract surplus instead, by pricing above cost and living with lower attendance. Which effect dominates in a given run is an empirical question about the balance between coordination gains and monopoly-style extraction, not something settled by the model's structure alone.

There is also a dynamic dimension: if the bar faces a binding sustainability constraint, strategic pricing may be necessary for long-run viability. Short-run monopoly distortions may be a worthwhile trade-off for long-run availability of the resource.

\section{Model evaluation} \label{sec: Model Evaluation}

The preceding sections describe co-evolution as a possibility: two learning processes coupled through price and attendance, which may settle down or may not, depending on how fast each side adapts. Under this section it is attested that possibility against simulated play across the full space of agent--bar pairings, and going beyond a single fixed horizon, across three different run lengths, so that claims about where the system ``ends up'' can be distinguished from claims about where it happens to be after an arbitrarily chosen number of rounds.

\subsection{Experimental design} \label{subsec: Experimental Design}

We simulate a population of $n = 10$ agents against a single bar, repeated across $10$ random seeds per condition. Four agent types are compared: a memory-based heuristic (the classical El Farol baseline, no learning beyond averaging recent observations), a Bayesian belief-updater, a no-regret learner (Multiplicative Weights over a small set of candidate attendance rules), and an individual Q-learner. Each is paired against four bar types: passive (fixed price), a regression-based demand forecaster, a Gaussian-process (GP) forecaster with an exploration bonus, and a Q-learning bar with constraint penalties built into its reward. This yields sixteen agent--bar pairings.

Unlike a single-horizon comparison, every pairing here is run to three separate horizons, $10$, $100$, and $1000$ rounds, with the seed-averaged mean and standard deviation of attendance $K_t$, price $p_t$, bar profit $\Pi_t$, and social welfare $W_t^{total}$ recorded at each horizon. The purpose of the added horizon axis is to ask a question the co-evolutionary framing of Section~\ref{sec: AI-Augmented Bar Intelligence & Co-Evolutionary Dynamics} raises but does not, on its own, answer: whether a given pairing's outcome at some arbitrary stopping point reflects a settled regime or an ongoing transient. Full implementation details and parameter values are given in the accompanying code repository in~\cite{AIFarolDataset2026}. 

All comparative statistics reported below are means and standard deviations over $10$ seeds per condition. Several cells, most notably no-regret/regression in Table~\ref{tbl: Full_Grid}, where the attendance standard deviation ($2.72$) exceeds its mean ($2.11$), are highly dispersed and plausibly skewed or multimodal across seeds. With only $10$ seeds, comparisons between pairings in Sunsections~\ref{subsec: The Steady-State Landscape}--\ref{subsec: Horizon Sensitivity: Transients, Not Equilibria} should be read as descriptive rather than as formally significance-tested differences; we have not computed confidence intervals or hypothesis tests on the gaps we describe as ``sharpest'' or largest, and a higher seed count would be needed to support such claims rigorously.

\subsection{The steady-state landscape} \label{subsec: The Steady-State Landscape}

\begin{figure}[t!]
	\centering
	\includegraphics[width=0.99\textwidth]{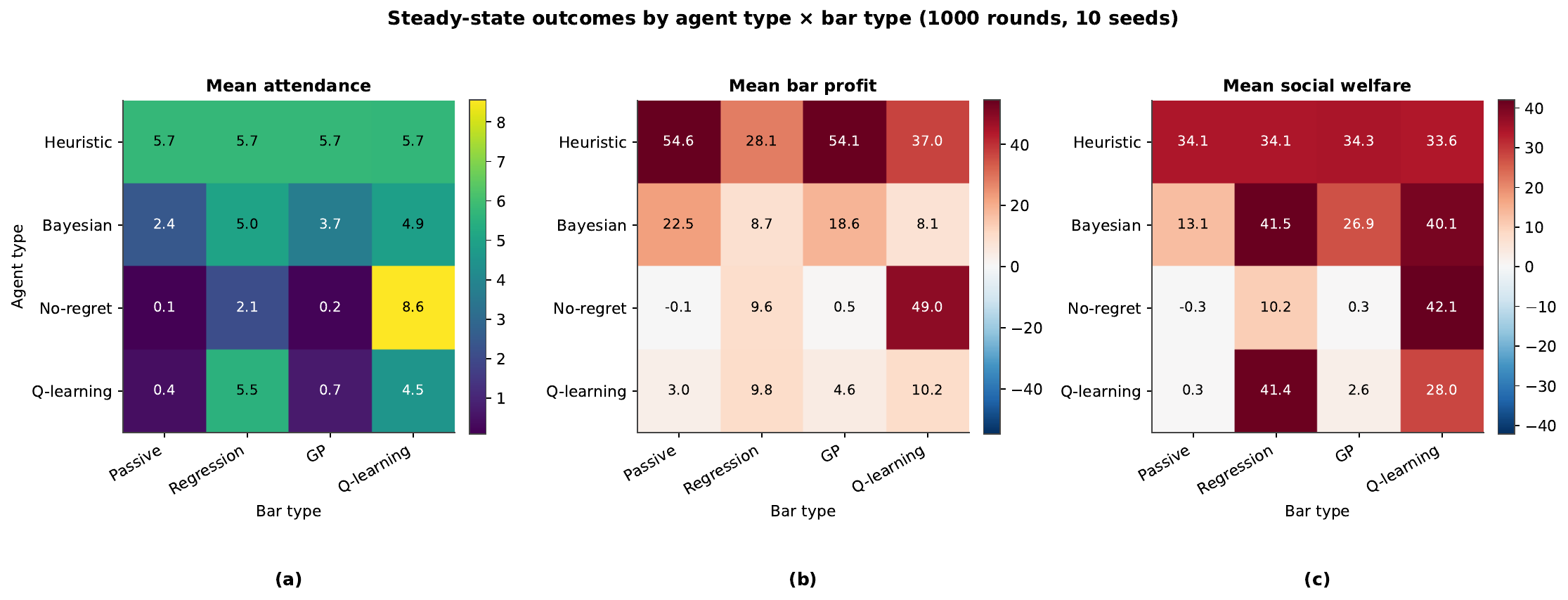}
	\caption{Steady-state outcomes for all sixteen agent--bar pairings at 1000 rounds, averaged over 10 seeds: (a) mean attendance; (b) mean bar profit; (c) mean social welfare. Attendance collapse under a non-adaptive bar is shared by both sophisticated agent types (no-regret, Q-learning) in (a); only the Q-learning bar substantially recovers attendance from either. Profit (b) and welfare (c) do not rank pairings the same way.}
	\label{fig: HeatMaps}
\end{figure}

Figure~\ref{fig: HeatMaps} summarizes mean attendance, bar profit, and social welfare for all sixteen pairings at the longest simulated horizon ($1000$ rounds), and Table~\ref{tbl: Full_Grid} reports the corresponding numerical values with cross-seed standard deviations. Two patterns are immediately visible that a smaller, hand-picked set of pairings would not reveal.

First, attendance collapse under a non-adaptive bar is not a peculiarity of no-regret learning. It also occurs, comparably severely, for individual Q-learning agents: mean attendance falls to $0.10$ (no-regret) and $0.42$ (Q-learning) against a passive bar, and to $0.16$ and $0.70$, respectively, against the GP forecaster. Both sophisticated agent types learn that, once congestion is priced into the reward signal, a policy that mostly stays home outperforms one that mostly attends whenever the bar does not adjust its price (or adjusts it in a way that does not track the agents' own responses). The mechanism described narratively in Section~\ref{sec: AI-Augmented Bar Intelligence & Co-Evolutionary Dynamics}, sophistication punishing passivity, is not specific to any one learning rule; the shared driver appears to be whether the agent revises attendance in response to realized crowding at all, rather than the particular algorithm used to do so.

Second, only the Q-learning bar reliably recovers attendance from sophisticated agents. Neither the regression nor the GP forecaster comes close: no-regret agents reach a mean attendance of $8.56$ against the Q-learning bar, versus $2.11$ and $0.16$ against the regression and GP bars respectively; individual Q-learning agents reach $4.50$ against the Q-learning bar, versus $5.53$ (regression) and $0.70$ (GP), the regression figure being a partial exception discussed in Subection~\ref{subsec: Variability & The Signature Of Persistent Adaptation}. Passive demand \emph{forecasting}, even a fairly capable one, is evidently not the same capability as active \emph{policy} learning: a bar that predicts demand accurately but does not adjust its objective in response to how its own pricing reshapes that demand does not, in this setup, out-perform a bar that does nothing at all.

Third, profit and welfare rank the sixteen pairings differently. The heuristic/passive pairing attains the single highest bar profit in the grid ($54.63$) but only the joint-sixth-highest welfare ($34.13$); the no-regret/Q-learning-bar pairing attains slightly lower profit ($48.96$) but the highest welfare in the grid ($42.12$); and the Bayesian/regression pairing attains the second-highest welfare ($41.50$) while its bar profit is among the lowest ($8.69$), a case in which the regression bar's pricing gives away most of the surplus it helps create rather than capturing it. Since Section~\ref{sec: Equilibrium Analysis & Welfare} defines welfare specifically to separate genuine coordination gains from one-sided surplus extraction, this divergence between the profit-maximizing and welfare-maximizing pairings is a directly relevant empirical result: a bar that is highly successful by its own objective is not automatically the one that best resolves the underlying coordination problem for the population as a whole.

\begin{table}[t!]
	\centering
	\caption{Steady-state outcomes (1000 rounds, mean over 10 seeds; standard deviation in parentheses).}
	\label{tbl: Full_Grid}
	\begin{tabular}{llrrrr}
		\toprule
		Agent type & Bar type & $K_t$ & $p_t$ & $\Pi_t$ & $W_t^{total}$ \\
		\midrule
		Heuristic  & Passive    & 5.74 (0.19) & 10.00 (0.00) & 54.63 (1.86)  & 34.13 (4.82)  \\
		Heuristic  & Regression & 5.74 (0.19) & 5.29 (1.64)  & 28.12 (10.84) & 34.13 (4.82)  \\
		Heuristic  & GP         & 5.74 (0.19) & 9.90 (0.36)  & 54.10 (3.22)  & 34.32 (4.33)  \\
		Heuristic  & Q-learning & 5.71 (0.20) & 6.85 (1.12)  & 36.99 (5.23)  & 33.58 (5.03)  \\
		Bayesian   & Passive    & 2.42 (0.21) & 10.00 (0.00) & 22.46 (2.03)  & 13.09 (2.09)  \\
		Bayesian   & Regression & 4.99 (0.05) & 2.32 (0.02)  & 8.69 (0.18)   & 41.50 (0.54)  \\
		Bayesian   & GP         & 3.67 (0.56) & 5.94 (1.94)  & 18.60 (3.67)  & 26.87 (6.44)  \\
		Bayesian   & Q-learning & 4.87 (0.11) & 2.62 (0.42)  & 8.10 (0.71)   & 40.05 (1.53)  \\
		No-regret  & Passive    & 0.10 (0.02) & 10.00 (0.00) & $-$0.08 (0.18) & $-$0.31 (0.17) \\
		No-regret  & Regression & 2.11 (2.72) & 3.28 (1.56)  & 9.56 (14.01)  & 10.24 (14.38) \\
		No-regret  & GP         & 0.16 (0.06) & 9.59 (0.19)  & 0.46 (0.49)   & 0.27 (0.48)   \\
		No-regret  & Q-learning & 8.56 (1.57) & 6.13 (0.60)  & 48.96 (9.18)  & 42.12 (8.30)  \\
		Q-learning & Passive    & 0.42 (0.03) & 10.00 (0.00) & 3.03 (0.25)   & 0.32 (0.12)   \\
		Q-learning & Regression & 5.53 (0.32) & 2.36 (0.26)  & 9.82 (1.97)   & 41.38 (4.26)  \\
		Q-learning & GP         & 0.70 (0.75) & 9.24 (1.18)  & 4.57 (4.51)   & 2.61 (5.99)   \\
		Q-learning & Q-learning & 4.50 (0.68) & 4.73 (0.84)  & 10.18 (2.55)  & 28.00 (7.73)  \\
		\bottomrule
	\end{tabular}
	\par\vspace{4pt}\noindent\footnotesize\textit{Note:} standard deviations are computed over $10$ seeds. Cells with a coefficient of variation above $1$ (notably No-regret/Regression, and to a lesser extent No-regret/Passive and Q-learning/GP) are highly dispersed and possibly skewed or multimodal across seeds; mean $\pm$ s.d.\ should be interpreted with this caveat for those rows, and reporting medians/interquartile ranges alongside the mean is recommended for these cells in future revisions.
\end{table}

\begin{figure}[t!]
	\centering
	\includegraphics[width=0.99\textwidth]{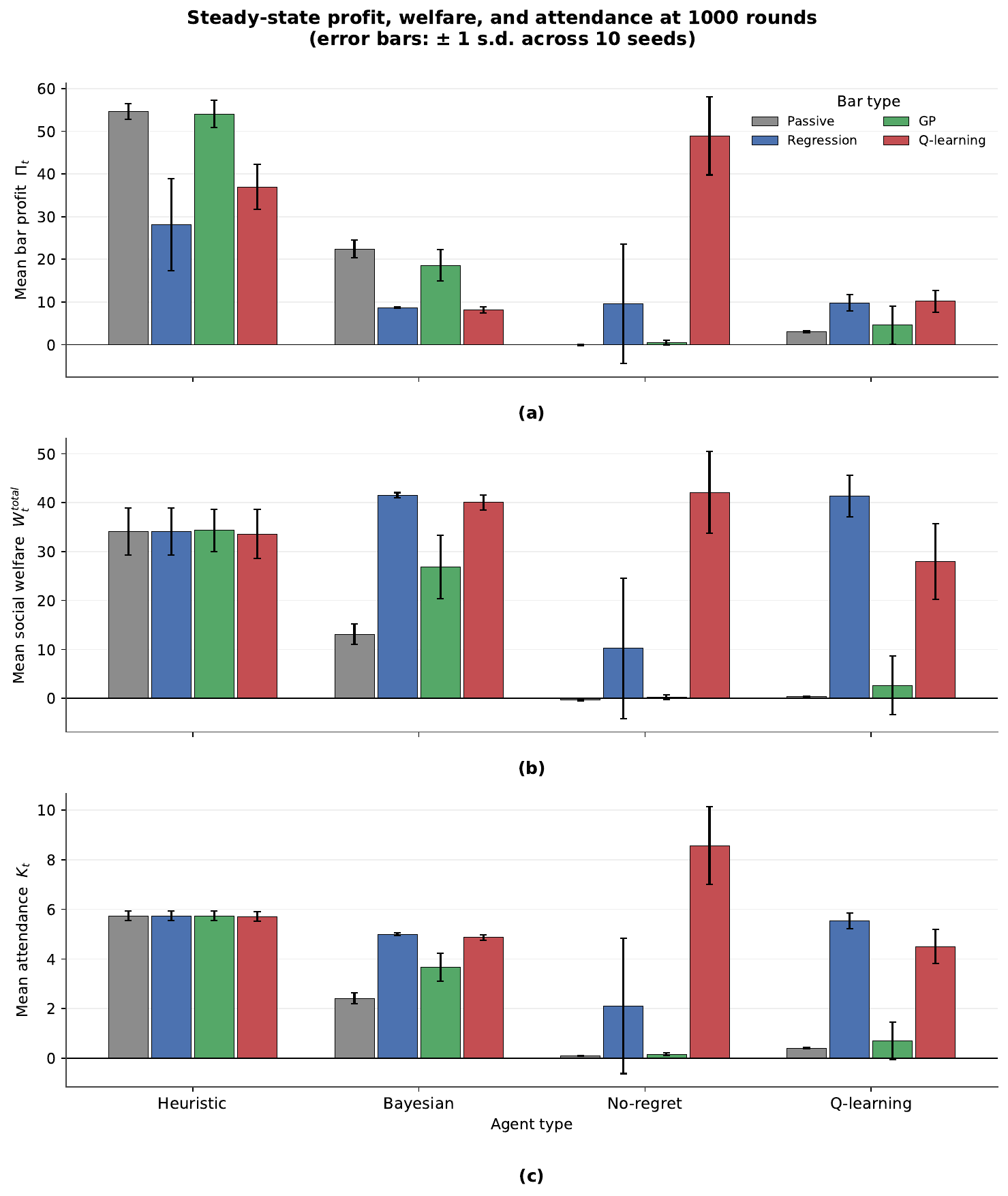}
	\caption{Steady-state outcomes at 1000 rounds for all sixteen pairings, grouped by agent type (error bars: $\pm 1$ standard deviation across 10 seeds): (a) mean bar profit; (b) mean social welfare; (c) mean attendance. Presented alongside Table~\ref{tbl: Full_Grid} to make the profit--welfare divergence discussed above, compare (a) and (b) for the same agent type, directly comparable across bar types within each agent type.}
	\label{fig: GroupedBars}
\end{figure}
\subsection{A passive Bar starves itself \& can go bankrupt doing it} \label{subsec: A Passive Bar Starves Itself & Can Go Bankrupt Doing It}

The clearest single finding in the grid concerns what happens when a sophisticated, learning population faces a bar that never moves. Table~\ref{tbl: Full_Grid}'s no-regret/passive cell shows mean attendance falling to $0.10$ by round $1000$, with mean bar profit turning slightly \emph{negative} ($-0.08$) and mean social welfare likewise negative ($-0.31$). Unlike the moderate-attendance, positive-profit version of this result reported at shorter horizons (see Section~\ref{subsec: Horizon Sensitivity: Transients, Not Equilibria}), the long-run outcome is a genuine market failure: the bar's fixed operating cost exceeds the revenue generated by the handful of customers who still show up, and the venue would, under this pricing rule, be running at a loss indefinitely. Nothing is broken in this outcome, the agents are correctly learning that attending rarely pays off once congestion and a fixed high price are both accounted for, and a bar that refuses to adjust its price prices itself out of its own market once its customers are sophisticated enough to notice, exactly as argued qualitatively in Section~\ref{sec: AI-Augmented Bar Intelligence & Co-Evolutionary Dynamics}. What the full grid adds is that this is not confined to no-regret learners: individual Q-learning agents facing a passive bar reach a similarly low steady-state attendance ($0.42$) and comparably poor welfare ($0.32$), though bar profit in that case stays weakly positive ($3.03$) rather than turning negative, a difference in degree rather than in kind between the two learning rules.

Pairing the same sophisticated agents with a learning bar reverses the picture, but not uniformly across bar types. Against the Q-learning bar, no-regret agents reach $8.56$ mean attendance and $48.96$ mean profit, the highest profit among the sophisticated-agent pairings in the grid, while individual Q-learning agents reach a more modest $4.50$ attendance and $10.18$ profit. Against the regression and GP forecasters, the picture is mixed rather than uniformly recovered: the regression bar does noticeably better against Q-learning agents ($5.53$ attendance) than against no-regret agents ($2.11$, with a very large cross-seed spread discussed below), while the GP bar fails to recover attendance from either sophisticated agent type ($0.16$ and $0.70$ respectively). Demand forecasting alone, even a comparatively expressive one such as a Gaussian process, is evidently not a substitute for the policy-learning objective that lets the Q-learning bar adapt its price to the type of customer actually attending.

\subsection{Horizon sensitivity: transients, not equilibria} \label{subsec: Horizon Sensitivity: Transients, Not Equilibria}

\begin{figure}[t!]
	\centering
	\includegraphics[width=0.99\textwidth]{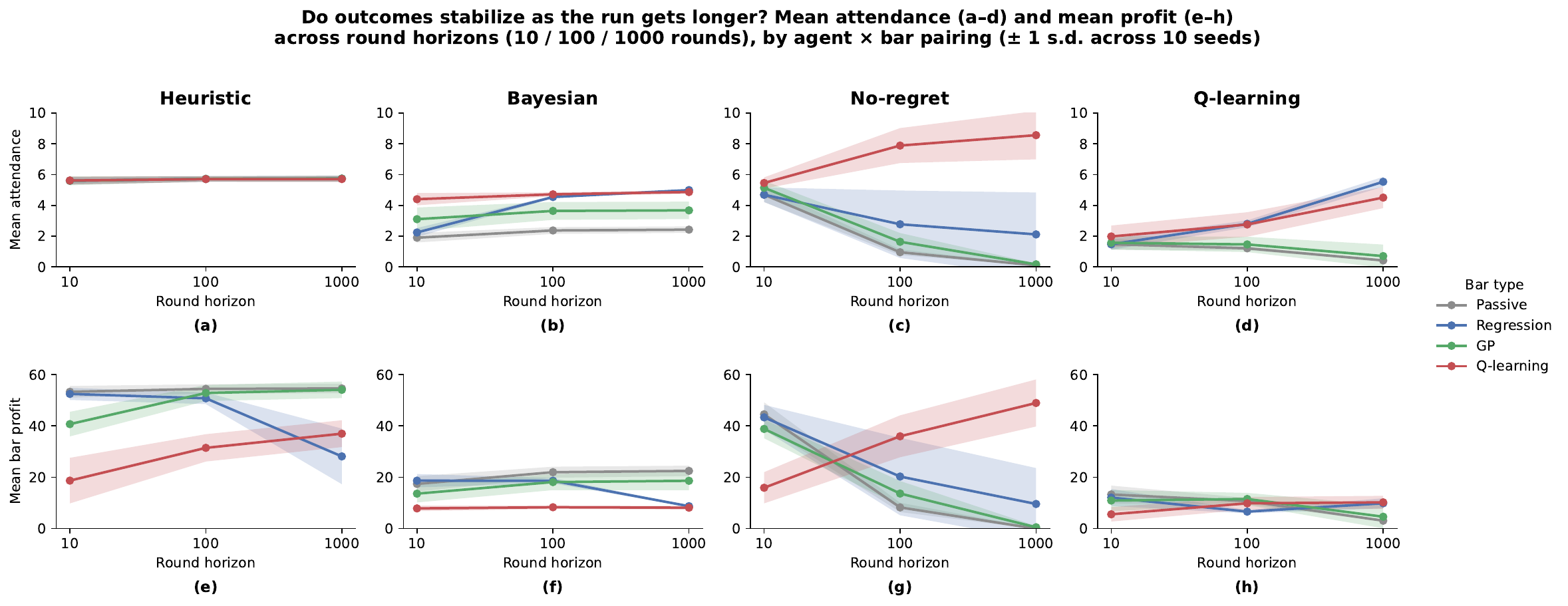}
	\caption{Mean attendance, (a) heuristic, (b) Bayesian, (c) no-regret, (d) Q-learning, and mean bar profit, (e) heuristic, (f) Bayesian, (g) no-regret, (h) Q-learning, as a function of round horizon, by bar type (colored lines), $\pm 1$ standard deviation across 10 seeds. Heuristic agents (a, e) are essentially flat across all three horizons; the other three agent types (b--d, f--h) show trajectories that have not visibly leveled off by round 1000 for at least one bar type each.}
	\label{fig: Convergence}
\end{figure}

A question the co-evolutionary framing in Section~\ref{sec: AI-Augmented Bar Intelligence & Co-Evolutionary Dynamics} raises but that a single-horizon comparison cannot answer is whether a reported outcome reflects a settled regime or a system still in motion. Figure~\ref{fig: Convergence} plots mean attendance and mean bar profit at each of the three simulated horizons, $10$, $100$, and $1000$ rounds, for every agent type, with lines for each bar type and shaded bands showing $\pm 1$ standard deviation across seeds.

Only the heuristic agent type is settled by the shortest horizon tested: attendance is flat at $5.60$--$5.74$ across all four bar types and all three horizons, reproducing Arthur's original result regardless of how sophisticated the bar's pricing algorithm is. Every other agent type shows at least one pairing whose trajectory has not flattened by round $1000$.

Against passive and GP bars, both no-regret and Q-learning agents show monotonic decline that does not visibly reach a floor within the simulated horizon: no-regret/passive attendance falls from $4.70$ ($10$ rounds) to $0.95$ ($100$ rounds) to $0.10$ ($1000$ rounds), with each order-of-magnitude increase in horizon producing roughly another five-fold reduction, a trajectory consistent with continued decay toward zero rather than convergence to a small positive attendance level. The no-regret/Q-learning-bar pairing moves in the opposite direction and is, if anything, further from settling: attendance rises from $5.46$ to $7.89$ to $8.56$ and profit from $15.9$ to $36.0$ to $49.0$ across the same three horizons, with no sign of the increments shrinking between the last two points.

The sharpest qualitative change in the grid belongs to Bayesian agents against the regression bar. Mean attendance nearly doubles across the three horizons ($2.24 \to 4.54 \to 4.99$) while price collapses from $9.24$ to $2.32$, and  unlike the no-regret/passive and no-regret/Q-learning-bar cases, where the cross-seed spread stays roughly constant or grows, the cross-seed standard deviation of attendance in this cell \emph{shrinks} by a factor of six over the same span ($0.29 \to 0.08 \to 0.05$). This is not merely a system still drifting; it is one converging tightly onto a qualitatively different regime from the one it started in, and it is arguably the most interesting single trajectory in the dataset.

\begin{figure}[t!]
	\centering
	\includegraphics[width=0.99\textwidth]{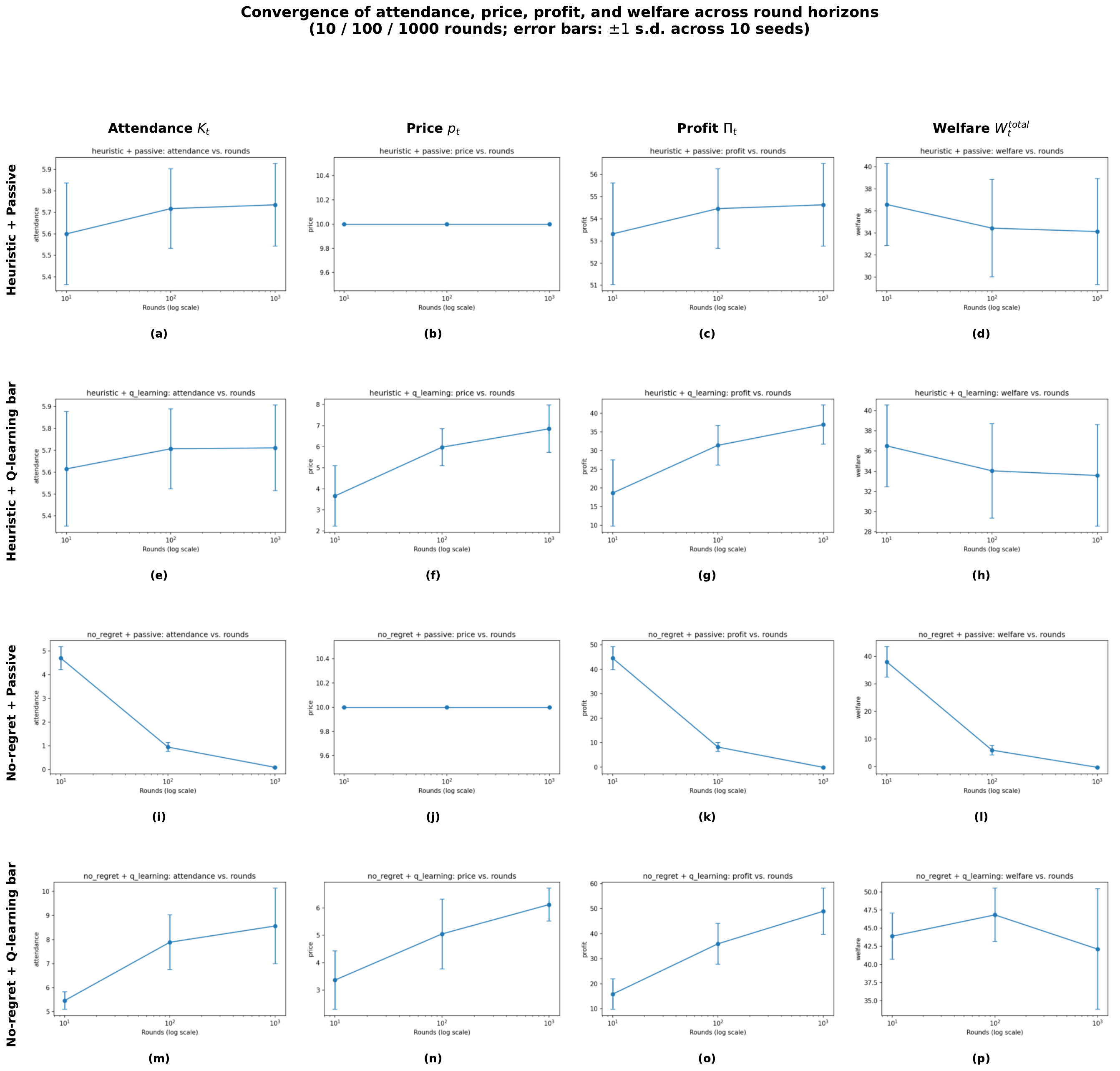}
	\caption{Attendance, price, profit, and welfare as a function of round horizon (10 / 100 / 1000 rounds; error bars: $\pm 1$ standard deviation across 10 seeds) for four representative pairings spanning the extremes of the grid. Heuristic + Passive: (a) attendance, (b) price, (c) profit, (d) welfare. Heuristic + Q-learning bar: (e) attendance, (f) price, (g) profit, (h) welfare. No-regret + Passive: (i) attendance, (j) price, (k) profit, (l) welfare. No-regret + Q-learning bar: (m) attendance, (n) price, (o) profit, (p) welfare.}
	\label{fig: FullConvergence}
\end{figure}

Figure~\ref{fig: Convergence} shows attendance and profit only; Figure~\ref{fig: FullConvergence} completes the picture for four representative pairings chosen to span the extremes of the grid, heuristic and no-regret agents, each against the passive and the Q-learning bar, by adding price and welfare trajectories alongside attendance and profit.

Two things emerge from the added price and welfare panels that Figure~\ref{fig: Convergence} alone does not show. First, welfare under the passive bar tracks attendance closely in both cases, as Section~\ref{sec: Equilibrium Analysis & Welfare} predicts it should, since price does not enter $W_t^{total}$: heuristic/passive welfare (d) stays essentially flat ($36.6 \to 34.4 \to 34.1$) alongside flat attendance (a), while no-regret/passive welfare (l) collapses in step with attendance (i), crossing into negative territory by round $1000$ ($37.9 \to 5.9 \to -0.3$). Second, welfare under the Q-learning bar does \emph{not} simply track the still-rising attendance and profit documented in Subsection~\ref{subsec: Horizon Sensitivity: Transients, Not Equilibria}: the no-regret/Q-learning-bar pairing's welfare (p) rises from $43.9$ (10 rounds) to a peak of $46.8$ (100 rounds) before falling back to $42.1$ (1000 rounds), even though attendance (m) and profit (o) are both still increasing monotonically at that same horizon. This non-monotonicity is consistent with the congestion term in $U(K_t)$ beginning to bind as attendance is pushed toward capacity: the bar's pricing keeps extracting more revenue and pulling in more customers, but past some point the marginal customer's own enjoyment, and hence aggregate welfare, starts to fall even as profit keeps climbing, a divergence between the bar's objective and total surplus that Table~\ref{tbl: Full_Grid} and Figure~\ref{fig: GroupedBars} could only show at a single horizon and that only becomes visible once welfare is tracked across the same horizon axis as profit.

The practical consequence for how these results should be read is that a reported outcome at a single, moderate horizon cannot be presumed representative of long-run behavior without checking the trend. At least four of the sixteen pairings examined here, no-regret/passive, no-regret/Q-learning-bar, Bayesian/regression, and Q-learning/regression, are qualitatively different at $1000$ rounds from what they look like at $10$ or $100$ rounds, not merely more precisely estimated versions of the same number.

\subsection{Variability \& the signature of persistent adaptation} \label{subsec: Variability & The Signature Of Persistent Adaptation}

Section~\ref{sec: AI-Augmented Bar Intelligence & Co-Evolutionary Dynamics} argues that fast mutual adaptation on both sides can sustain oscillation or limit-cycle behavior rather than settling to a fixed point, while slow adaptation tends toward an approximate equilibrium. The round-level data needed to test this directly, an auto-correlation or spectral analysis of $K_t$ within a single long run, is a finer-grained question than the seed-averaged summary statistics analyzed here can answer on their own. Short of that, the dispersion of outcomes \emph{across} independently-seeded runs, and how that dispersion changes as each run gets longer, offers an indirect but informative proxy: for a process converging to a stable fixed point, longer runs should produce more consistent (lower-variance) outcomes across seeds, whereas a process with a persistent limit cycle or strong path-dependence can continue to show substantial or even growing across-seed dispersion regardless of run length, because different seeds may settle into different phases of the cycle or different basins entirely.

\begin{figure}[t!]
	\centering
	\includegraphics[width=0.99\textwidth]{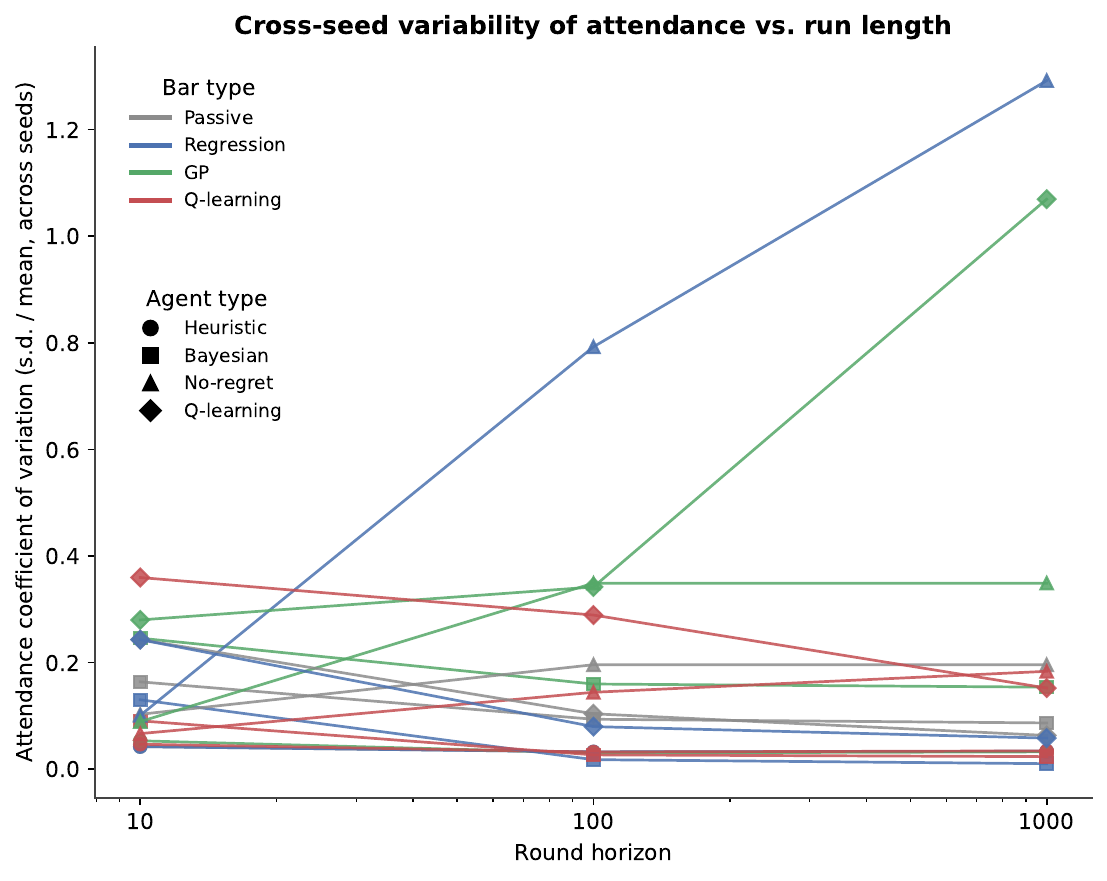}
	\caption{Coefficient of variation (cross-seed standard deviation divided by the mean) of attendance, plotted against round horizon, for all sixteen pairings. Most pairings' dispersion falls or stays flat as the horizon grows, consistent with convergence to a stable outcome. The no-regret/Q-learning-bar pairing is an exception: its coefficient of variation rises monotonically across all three horizons.}
	\label{fig: Variability}
\end{figure}

Figure~\ref{fig: Variability} shows that most pairings' attendance dispersion falls or stays flat as the horizon lengthens, the ordinary signature of averaging over a longer, more representative sample of a process that is settling down, and consistent with the tight convergence documented for Bayesian/regression in Subsection~\ref{subsec: Horizon Sensitivity: Transients, Not Equilibria}. The no-regret/Q-learning-bar pairing is the clear exception: its coefficient of variation rises across all three horizons ($0.067 \to 0.144 \to 0.183$), even as the run itself grows longer. Rising across-seed dispersion under a longer observation window, for a pairing whose mean attendance and profit are simultaneously still increasing (Subection~\ref{subsec: Horizon Sensitivity: Transients, Not Equilibria}), is consistent with, though it does not on its own prove, the persistent-adaptation regime described qualitatively in Section~\ref{sec: AI-Augmented Bar Intelligence & Co-Evolutionary Dynamics}, in which neither learner ever fully settles and different seeds can end up on different points of an ongoing trajectory. Confirming this properly would require examining $K_t$ within a single run at fine time resolution, which is a natural next step given the present result.

A second, unrelated instability is worth flagging separately: the regression bar's price standard deviation against \emph{heuristic} agents grows nearly thirty-fold between the $100$- and $1000$-round horizons ($0.18 \to 1.64$), with profit standard deviation growing correspondingly ($2.3 \to 10.8$ against a mean of $28.1$), even though heuristic attendance itself barely moves and its own variance stays tight ($5.60 \to 5.72 \to 5.74$, standard deviation consistently under $0.2$). Since the growing dispersion originates entirely on the bar's side while the customer population is essentially static, this looks less like co-evolutionary instability and more like a demand-model misspecification problem: the regression forecaster appears to mistake ordinary sampling noise in a nearly price-insensitive population for a genuine demand signal, and different seeds' regression fits diverge onto different, self-reinforcing pricing trajectories as a result. This is a concrete instance of the point made qualitatively in Section~\ref{sec: AI-Augmented Bar Intelligence & Co-Evolutionary Dynamics}, that the bar's forecasting method, not just its objective, shapes the outcome, and suggests that not every source of instability observed in this framework reflects genuine two-sided co-evolution; some of it can arise from one side mismodeling a population that is not, in fact, adapting.

\subsection{Checking the no-regret claim} \label{subsec: Checking The No-Regret Claim}

Section~\ref{sec: AI-Augmented Bar Intelligence & Co-Evolutionary Dynamics} cites Hart and Mas-Colell's result~\cite{Hart2000} that no-regret learning drives play toward a coarse correlated equilibrium, and Section~\ref{sec: Equilibrium Analysis & Welfare} leans on that same result to justify CCE as the relevant solution concept once agents are boundedly rational learners rather than equilibrium calculators. That is a claim about what happens in the limit; the horizon-sensitivity results of Subsection~\ref{subsec: Horizon Sensitivity: Transients, Not Equilibria} give a further reason to check it directly rather than take it on faith, since several pairings in this study are still transitioning at the longest horizon tested.

\begin{figure}[t!]
	\centering
	\includegraphics[width=0.99\textwidth]{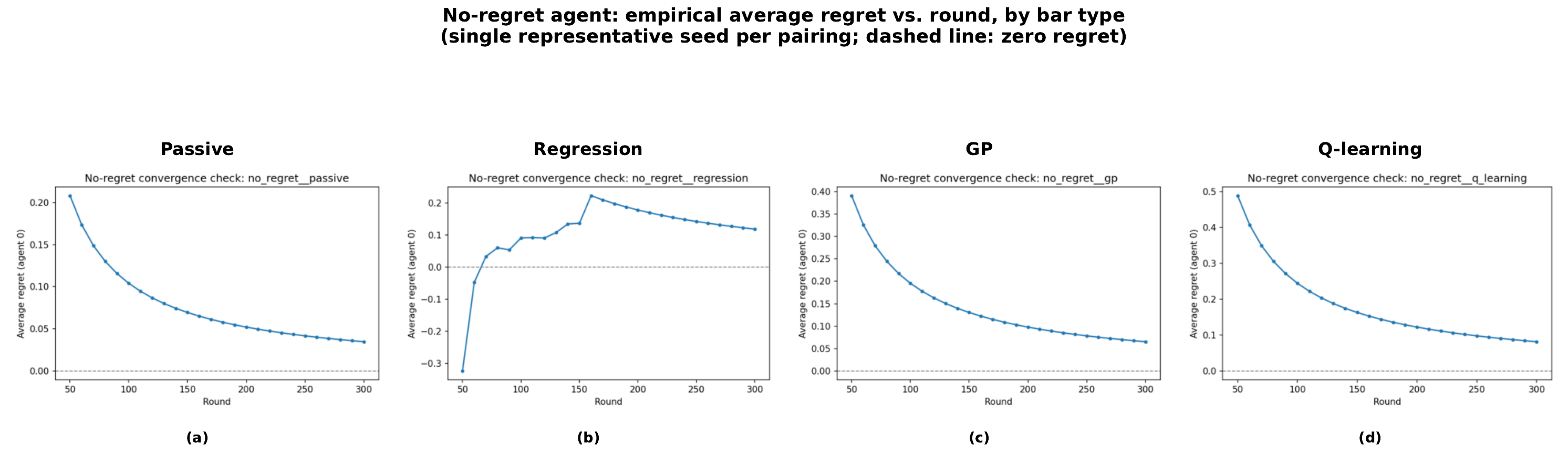}
	\caption{Empirical average regret of the no-regret agent, rounds 50--300 of a single representative run, against each of the four bar types (dashed line: zero regret): (a) passive; (b) regression; (c) GP; (d) Q-learning. Regret decays smoothly toward zero in (a), (c), and (d). In (b), regret is initially negative, rises and crosses zero by round $\sim 65$, peaks around round $160$, and only then begins to decay, a qualitatively different trajectory from panels (a), (c), and (d). Note that the no-regret guarantee bounds the gap to the best \emph{fixed} strategy in hindsight, not the level at which the system's aggregate statistics eventually settle; vanishing regret in panel (d) is therefore consistent with, not contradicted by, the still-rising mean attendance and profit reported for that pairing in Subsection~\ref{subsec: Horizon Sensitivity: Transients, Not Equilibria}.}
	\label{fig:regret}
\end{figure}

For the no-regret agent, we can compute, after the fact, using the realized sequence of attendance and price, what that agent would have earned had it played each of its candidate strategies exclusively throughout the run, and compare that to what it actually earned. The gap between the best such fixed strategy and the agent's realized payoff, averaged per round, is the empirical regret; the no-regret guarantee says this quantity should shrink toward zero as the run gets longer. Figure~\ref{fig:regret} plots this empirical regret between round $50$ and round $300$ of a single representative run, separately for the no-regret agent against each of the four bar types.

Against three of the four bar types, passive (a), GP (c), and Q-learning (d), the pattern is exactly what the no-regret guarantee predicts: regret starts appreciably above zero (roughly $0.21$, $0.39$, and $0.49$ respectively at round $50$) and decays smoothly and monotonically toward it, reaching roughly $0.03$--$0.08$ by round $300$. The Q-learning-bar panel (d) is the noisiest of the three but still shows the same qualitative shape as (a) and (c), despite that pairing being the one whose \emph{mean} attendance and profit were still visibly rising at round $1000$ in Subsection~\ref{subsec: Horizon Sensitivity: Transients, Not Equilibria}, regret decay and non-stationary mean outcomes are not mutually exclusive, since a no-regret guarantee only bounds the gap to the best fixed strategy in hindsight, not the level the system's aggregate statistics eventually reach.

Against the regression bar (b), the picture is qualitatively different, and consistent with the instability already flagged for this pairing in Subsection~\ref{subsec: Variability & The Signature Of Persistent Adaptation} (extreme cross-seed spread in attendance and price at long horizons). Regret starts \emph{negative} at round $50$ (around $-0.32$, meaning the agent was, up to that point, doing better than its own best fixed candidate strategy in hindsight), rises sharply and crosses zero by roughly round $65$, continues climbing to a peak of about $0.22$ around round $160$, and only then turns over and begins to decay, reaching roughly $0.12$ by round $300$, still well above where panels (a), (c), and (d) are at the same round. This non-monotonic shape is not consistent with the clean vanishing-regret trajectory the other three pairings display, at least not within the window examined here. A plausible explanation is that the regression bar's own pricing behavior against this agent type is itself unstable (Subsection~\ref{subsec: Variability & The Signature Of Persistent Adaptation}), which means the no-regret agent is chasing a moving target whose statistics have not settled, the regret bound still applies asymptotically, but convergence can be substantially slower, and non-monotonic along the way, when the environment the agent is learning against is itself still adapting unpredictably.

Two conclusions follow. First, the CCE-convergence story motivating Section~\ref{sec: Equilibrium Analysis & Welfare} is not merely a theorem sitting apart from the model, its signature, decaying average regret, is directly visible in simulation for three of the four bar types, which is a meaningfully stronger empirical grounding for that solution concept than the aggregate outcome statistics alone provide. Second, the exception is informative rather than a nuisance: it shows that the no-regret guarantee's asymptotic character matters in practice, and that pairing a learning agent with a bar whose own policy is not yet stable can produce transient regret dynamics that look nothing like the textbook decay curve, even though both learners are behaving exactly as designed.

\subsection{Revealed pricing behavior} \label{subsec: Revealed Pricing Behavior}

\begin{figure}[t!]
	\centering
	\includegraphics[width=0.99\textwidth]{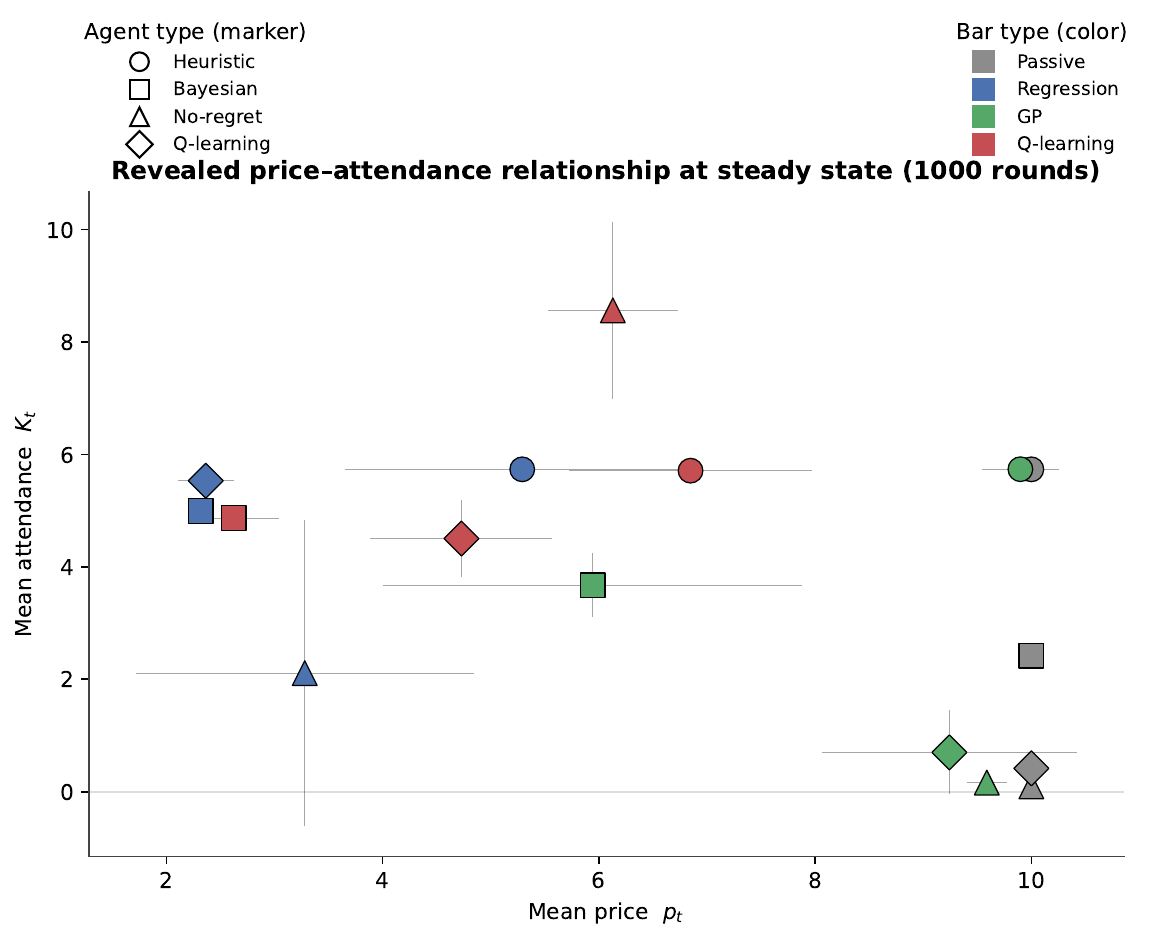}
	\caption{Steady-state price against steady-state attendance for all sixteen pairings at 1000 rounds (error bars: $\pm 1$ standard deviation across seeds; color denotes bar type, marker denotes agent type).}
	\label{fig:price_attendance}
\end{figure}

Figure~\ref{fig:price_attendance} plots each pairing's steady-state price against its steady-state attendance, making the qualitative distinction between bar types visible in a single view.

Passive-bar outcomes lie, by construction, on the vertical line $p_t = 10$, and span nearly the full observed range of attendance depending entirely on which agent type is on the other side, a direct illustration of the Subsection~\ref{subsec: A Passive Bar Starves Itself & Can Go Bankrupt Doing It} point that a fixed price leaves the coordination outcome entirely up to customer sophistication. Regression- and GP-bar outcomes cluster toward higher prices with low-to-moderate attendance, with the regression bar's two low-price, higher-attendance points (against Bayesian and Q-learning agents) as partial exceptions. The Q-learning bar is the only type that reaches both a high-price/high-attendance point (against heuristic agents, price $6.85$, attendance $5.71$) and a comparatively low-price, high-attendance point (against no-regret agents, price $6.13$, attendance $8.56$), that is, the only bar policy in the grid that adapts its price differently enough across customer types to land in different regions of the price--attendance plane depending on who it is facing, rather than converging toward a single characteristic operating point regardless of the customer population.

\subsection{Discussion of the results} \label{subsec: Discussion Of The Results}

Six findings stand out from the full agent--bar--horizon grid. First, a bar's willingness to adapt matters far more in the presence of sophisticated, learning customers than with naive ones, passivity is not merely survivable but essentially costless against heuristic agents, and is self-defeating, up to and including negative long-run profit and welfare, against both no-regret and individual Q-learning agents. Second, this finding generalizes across learning-rule families rather than being a peculiarity of no-regret learning specifically. Third, several pairings are still visibly transitioning at $1000$ rounds rather than having reached a steady state, which means single-horizon comparisons, including the representative pairings highlighted in Subsections~\ref{subsec: The Steady-State Landscape}--\ref{subsec: A Passive Bar Starves Itself & Can Go Bankrupt Doing It}, should be read as snapshots along a trajectory rather than as terminal outcomes unless the horizon sensitivity has been checked, as it is here in Subsection~\ref{subsec: Horizon Sensitivity: Transients, Not Equilibria}. Fourth, the growing cross-seed dispersion of the no-regret/Q-learning-bar pairing is consistent with the persistent-adaptation regime the model anticipates when both sides learn quickly, though confirming it rigorously requires round-level analysis beyond what the present seed-aggregated data can provide; a separate, unrelated instability traced to the regression bar's demand model against a static customer population illustrates that not all observed instability in this framework is co-evolutionary in origin. Fifth, profit and welfare are only loosely aligned across the grid: the profit-maximizing pairing for the bar is not the welfare-maximizing pairing for the system as a whole, which matters directly for the mechanism-design claims motivating this paper, since a bar optimizing its own objective is not thereby guaranteed to be resolving the underlying coordination problem efficiently. Sixth, the no-regret guarantee underlying the CCE solution concept of Section~\ref{sec: Equilibrium Analysis & Welfare} is directly visible in simulation, decaying average regret against three of the four bar types, but breaks down into a slower, non-monotonic trajectory against the one bar type (regression) already flagged as unstable on other grounds, underscoring that the asymptotic character of the no-regret bound can matter substantially within any horizon actually simulated.

\section{Conclusions} \label{sec: Conclusions}

This section interprets what the results imply for the broader modeling framework: what changes, conceptually, once the venue is treated as a learner rather than a fixed rule; where the framework could be usefully applied and what its current formulation leaves out; and which extensions the simulation evidence itself points to most directly, as distinct from open problems that follow from the model alone.

\subsection{Remarks} \label{subsec: Remarks}

Treating the bar as a learner rather than a fixed rule is not merely a modeling nicety: it changes which equilibrium concepts apply, how welfare should be measured, and whether the system settles down at all. The simulation grid substantiates each part of that claim rather than leaving it at the level of narrative argument. Subsection~\ref{subsec: A Passive Bar Starves Itself & Can Go Bankrupt Doing It} shows that a fixed price can be self-defeating, up to and including negative long-run profit, once customers are sophisticated enough to notice; Subsection~\ref{subsec: The Steady-State Landscape} shows that the pairing that is best for the bar's own objective is not the pairing that is best for the system as a whole; and Subsections~\ref{subsec: Horizon Sensitivity: Transients, Not Equilibria}--\ref{subsec: Checking The No-Regret Claim} show that whether a reported outcome reflects a settled regime, an ongoing transient, or a genuine limit cycle cannot be read off a single-horizon snapshot, and must instead be checked against horizon sensitivity, cross-seed dispersion, and empirical regret directly. Questions of this kind,  what an adaptive institution should optimize for, and how its success should be judged relative to the population it serves rather than only against its own objective, extend well beyond the single bar studied here to any adaptive institution facing a learning user base.

\subsection{Potentials \& limitations} \label{subsec: Potentials & Limitations}

The practical reach of the two-sided learning framework and the assumptions that currently bound it are best read together, since the applications below inherit the same simplifications flagged as limitations. The framework developed here applies to numerous real-world settings: AI-driven revenue management in hospitality~\cite{Kastius2022, Bondoux2020}; spot pricing on cloud computing platforms balancing utilization and revenue~\cite{Kumar2018}; surge pricing in ride-sharing where both drivers and passengers learn and adapt~\cite{Garg2021}; and congestion pricing in communication networks~\cite{Henderson2001}. In all these cases, the co-evolutionary dynamic between an adaptive institution and a learning user base is the central feature, and the qualitative findings of Section~\ref{sec: Model Evaluation}, that passivity is costly against sophisticated users, and that a profit-optimal policy need not be a welfare-optimal one, plausibly transfer to these settings even though the present paper studies only a single, stylized venue.

On the limitation of the proposed approach is that the model is stylized to a single bar with homogeneous customers and a single pricing instrument. Real markets feature heterogeneous customers, multiple competing venues, and richer strategy spaces. The companion paper~\cite{Polenakis2026} explores incentive design under partial observability in a more constrained setting; extending those results to the full two-sided learning framework remains an open problem. Rigorous convergence analysis remains open: establishing the conditions under which customer learning and bar policy learning jointly settle down, and identifying which parameter choices keep the system stable, would call on tools from stochastic approximation and dynamical systems that fall outside the scope of the present paper. The simulation evidence itself is also bounded in scope, $n=10$ agents, a single bar, and ten seeds per condition (Section~\ref{sec: Model Evaluation}), and while the qualitative patterns we report are unlikely to be artifacts of these specific choices, the precise numerical thresholds at which, for instance, attendance collapse or profit--welfare divergence occurs have not been shown to be robust to population size or seed count, and should be treated as illustrative rather than as calibrated estimates.

\subsection{Future research} \label{subsec: Future Research}

Two concrete next steps follow directly from the simulation results in Section~\ref{sec: Model Evaluation} rather than from the model alone. First, the claim in Section~\ref{sec: AI-Augmented Bar Intelligence & Co-Evolutionary Dynamics} that oscillation versus convergence is governed by each side's learning rate is argued qualitatively and supported only indirectly here, via the cross-seed dispersion diagnostic of Subsection~\ref{subsec: Variability & The Signature Of Persistent Adaptation}; the present grid varies agent type and bar type, not learning-rate parameters, so the predicted transition from convergence to persistent oscillation as adaptation speed increases has not been directly demonstrated. A learning-rate sweep, holding agent and bar type fixed and varying step sizes on both sides, would test this central dynamical claim much more directly, and the same seed-dispersion and empirical-regret diagnostics used in Subsections~\ref{subsec: Variability & The Signature Of Persistent Adaptation}--\ref{subsec: Checking The No-Regret Claim} could be reused for it. Second, Subsection~\ref{subsec: Variability & The Signature Of Persistent Adaptation} shows that at least one instability in the grid (the regression bar's growing price variance against static heuristic agents) is attributable to demand-model misspecification rather than genuine two-sided co-evolution; disentangling this from the co-evolutionary instabilities emphasized elsewhere in the paper would benefit from a controlled ablation in which the regression bar is instead given the correctly specified functional form for demand, checking directly whether the instability disappears against a non-adapting population.

Further open questions include: how closely do learned policies approximate analytically optimal mechanisms; how can fairness constraints be incorporated into the bar's optimization; and how do customers strategically manipulate the bar's learning algorithm if they understand its structure? A further, more basic extension is to relax the single-agent-population, single-venue assumption flagged in Subsection~\ref{subsec: Potentials & Limitations}, heterogeneous customer segments and multiple competing venues would let the framework speak directly to market-level rather than single-venue outcomes. Finally, the model generates testable predictions about the co-evolution of pricing and attendance that warrant empirical validation against data from restaurants, bars, or ride-sharing platforms.

\bibliographystyle{ieeetr}
\bibliography{AI-Farol}

\end{document}